\RequirePackage{lineno}
\documentclass[aps,prb,10pt,twocolumn,superscriptaddress,longbibliography]{revtex4-2}

\usepackage{graphicx}
\usepackage{amssymb,amsmath,amsfonts,bm,hyperref}
\usepackage{color}
\usepackage{ulem}
\usepackage{bbm}
\usepackage{graphicx}
\usepackage{dcolumn}
\usepackage{mathtools}
\usepackage{pifont}
\usepackage{diagbox}

\usepackage{extarrows}
\usepackage{amsthm}
\usepackage{mathrsfs}
\usepackage{blindtext}
\usepackage{titlesec}
\usepackage{multirow}
\usepackage{bigstrut}

\def\I{\uppercase\expandafter{\romannumeral 1}}
\def\II{\uppercase\expandafter{\romannumeral 2}}
\def\III{{\uppercase\expandafter{\romannumeral 3}}}
\def\IV{{\uppercase\expandafter{\romannumeral 4}}}
\def\V{{\uppercase\expandafter{\romannumeral 5}}}
\def\VI{{\uppercase\expandafter{\romannumeral 6}}}
\def\VII{{\uppercase\expandafter{\romannumeral 7}}}
\def\VIII{{\uppercase\expandafter{\romannumeral 8}}}
\def\i{\lowercase\expandafter{\romannumeral 1}}
\def\ii{\lowercase\expandafter{\romannumeral 2}}
\def\iii{{\lowercase\expandafter{\romannumeral 3}}}
\def\iv{{\lowercase\expandafter{\romannumeral 4}}}
\def\v{{\lowercase\expandafter{\romannumeral 5}}}
\def\vi{{\lowercase\expandafter{\romannumeral 6}}}
\def\vii{{\lowercase\expandafter{\romannumeral 7}}}
\def\viii{{\lowercase\expandafter{\romannumeral 8}}}
\def\nn{\nonumber\\}

\def\k{\textbf{k}}

\def\M{\textbf{M}}
\def\m{\textbf{m}}
\def\vr{\textbf{r}}

\def\nn{\nonumber\\}

\def\L{\textrm{L}}

\def\RR{\textbf{R}}

\newcommand{\ket}[1]{\vert #1 \rangle}
\newcommand{\bra}[1]{\langle #1 \vert}
\newcommand{\braket}[2]{\langle #1 \vert #2 \rangle}

\begin{document}
\let\oldequation\equation
\let\oldendequation\endequation
\renewenvironment{equation}{\linenomathNonumbers\oldequation}{\oldendequation\endlinenomath}
%\linenumbers

\title{Orbital magnetoelectric coupling of three dimensional Chern insulators}

\author{Xin Lu}
\thanks{These authors contributed equally.}
\affiliation{School of Physical Science and Technology, ShanghaiTech University, Shanghai 201210, China}

\author{Renwen Jiang}
\thanks{These authors contributed equally.}
\affiliation{School of Physical Science and Technology, ShanghaiTech University, Shanghai 201210, China}

\author{Zhongqing Guo}
\affiliation{School of Physical Science and Technology, ShanghaiTech University, Shanghai 201210, China}

\author{Jianpeng Liu}
\email{liujp@shanghaitech.edu.cn}
\affiliation{School of Physical Science and Technology, ShanghaiTech University, Shanghai 201210, China}
\affiliation{ShanghaiTech Laboratory for Topological Physics, ShanghaiTech University, Shanghai 201210, China}
\affiliation{Liaoning Academy of Materials, Shenyang 110167, China}

\bibliographystyle{apsrev4-2}

\begin{abstract} 
Orbital magnetoelectric effect is closely related to the band topology of bulk crystalline insulators. Typical examples include the half quantized Chern-Simons orbital magnetoelectric coupling in three dimensional (3D) axion insulators and topological insulators, which are the hallmarks of their nontrivial bulk band topology. While the Chern-Simons coupling is well defined only for insulators with zero Chern number, the orbital magnetoelectric effects in 3D Chern insulators with nonzero (layer) Chern numbers are still open questions. In this work, we propose a never-mentioned quantization rule for the orbital magnetoelectric response in 3D Chern insulators, the spatial gradient of which is exactly quantized in unit of $e^2/h$. By theoretical analysis and numerical simulations, we demonstrate that such quantized orbital magnetoelectric response is exact for various types of interlayer hoppings and stackings, and remains robust even against disorder and lack of crystalline symmetries. We argue that the exact quantization has a topological origin and is protected by Chern number. Furthermore, we propose  two promising material platforms to observe the proposed quantized orbital magnetoelectric response thanks to recent experimental developments in detecting spatial magnetic-field distributions in  device systems.
\end{abstract} 

\maketitle

\section{Introduction}

Band insulators can be classified into different topological phases according to different topological invariants \cite{haldane-prl88, tknn, Mele-qsh-prl05, Mele-Z2-prl05}. One important and basic topological invariant is the Chern number for two dimensional (2D) systems \cite{tknn, kohmoto-85,haldane-prl88}, which describes the winding number of hybrid Wannier centers within 2D Brillouin zone \cite{vanderbilt-berry-book-2018,bernevig-topo-book-2013}. Those time-reversal-broken band insulators with nonzero Chern number are called Chern insulators, which would exhibit quantum anomalous Hall effect \cite{xue-qah-science13,zhang-qah-science20,young-qah-tbg-science20}. The notion of 2D Chern insulators can be generalized to its 3D counterpart by stacking an infinite number of 2D Chern insulators along the vertical axis \cite{halperin-3dqhe-jjaps-1987,hasan-ti-rmp-2010,qi-ti-rmp-2011,jin-3dqahe-prb-2018,wieder-weylcdw-prr-2020}. In principle, one can stack 2D Chern insulators along any direction so that one can define three independent Chern numbers $(C_x, C_y, C_z)$, called Chern vector \cite{kohmoto-3DQHE-pbcm-1993,haldane-3DAHE-prl-2004}, to characterize 3D band insulators. If the interlayer hopping is weak enough such that the bulk gap remains opened, such a topologically non-trivial 3D band insulator, named 3D Chern insulator, is adiabatically connected to an infinite number of decoupled 2D Chern-insulator layers. 

Topological 3D band insulators are known to exhibit non-trivial orbital magnetoelectric (OME) responses. The OME coupling $\alpha_{\mu\nu}$ is defined as the change in bulk electric polarization $P_\mu$ while applying magnetic field $B_\nu$, or equivalently the variation of bulk orbital magnetization $M_\nu$ induced by external electric field $\mathcal{E}_\mu$: $\alpha_{\mu\nu} = \partial P_\mu/\partial B_\nu = \partial M_\nu/\partial \mathcal{E}_\mu$ with $\mu,\nu=x, y, z$. This response is odd under inversion and time-reversal symmetry so that it vanishes if either of two symmetries is present. 3D time-reversal-protected topological insulators are classified by $\mathbb{Z}_2$ invariant \cite{Mele-Z2-prl05, Balents-Z2-prb07, Roy-qsh3D-prb09, Kane-Z2pump-prb06}, which exhibit a half quantized bulk diagonal Chern-Simons OME response $\alpha_{\mu\nu}=e^2/2h\,\delta_{\mu\nu}$, modulo an integer multiple of $e^2/h$, once the surface state is gapped out \cite{qi-topofield-prb08,chern-simons-prl09}.  %\Red{Usually, the Chern-Simons OME coupling is characterized by a dimensionless phase angle $\theta$, with $\alpha_{\mu\nu}=(\theta/2\pi)e^2/h\,\delta_{\mu\nu}$. Clearly, $\theta=0$ or $\pi$ in the presence of time-reversal or inversion symmetry exactly corresponds to the $\mathbb{Z}_2$ classification of 3D topological insulators or axion insulators \cite{qi-topofield-prb08,chern-simons-prl09}.}
However, the Chern-Simons coupling is well defined only for bulk band insulators with zero Chern number within any 2D plane in momentum space, namely with vanishing Chern vector. This is because the expression of the Chern-Simons OME coupling coefficient involves an integration of Chern-Simons 3-form defined by the occupied Bloch states over the 3D Brillouin zone \cite{chern-simons-prl09,qi-topofield-prb08,essin-ome-prb10,andrei-ome-njop11,coh-ome-prb11}, which requires the existence of a smooth and periodic gauge throughout the 3D Brillouin zone \cite{exponential-wannier-prl07,ti-smoothgauge-prb12,liu-gaugeCS-prb-2015}. However, such a gauge is non-existing for 3D Chern insulators due to topological obstruction. 

\begin{figure*}
    \centering
    \includegraphics[width=0.8\textwidth]{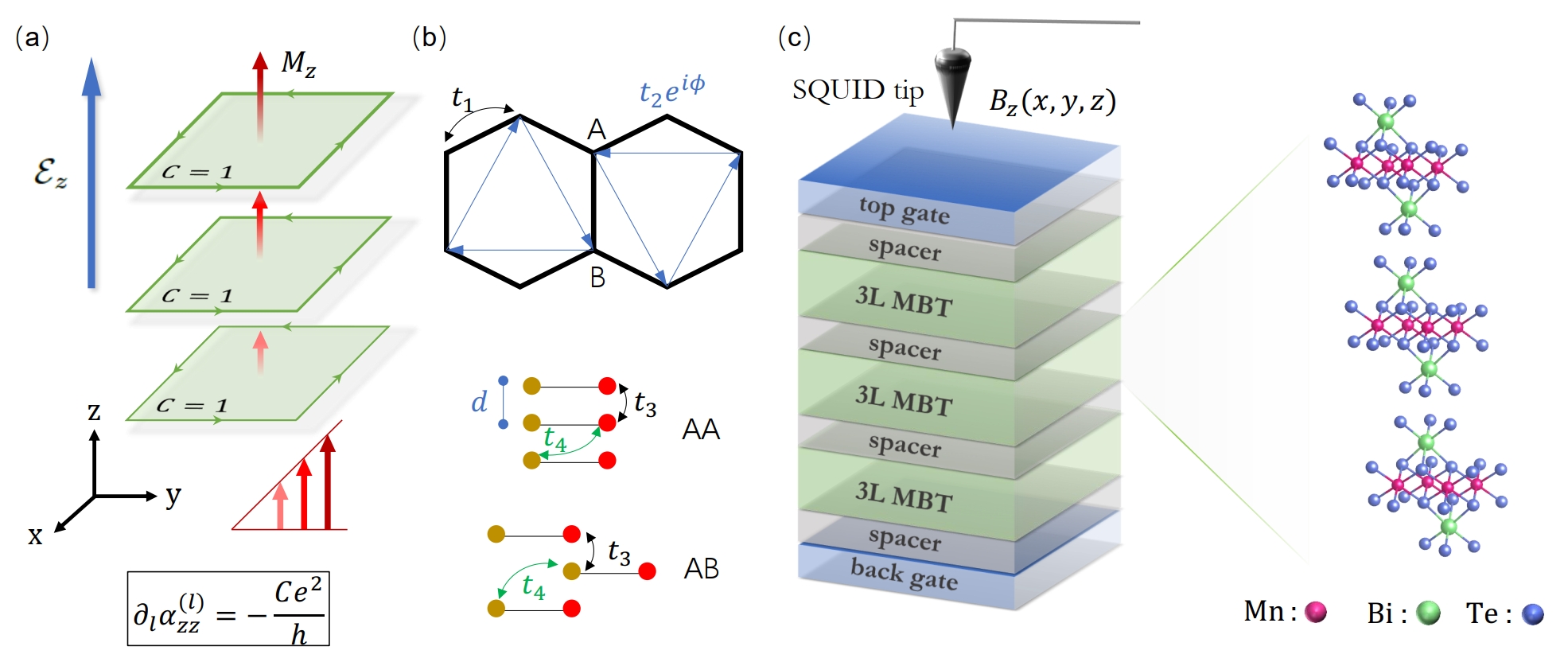}
    \caption{(a) Schematic illustration for the origin of the quantization rule Eq.~\eqref{eq:quantization_rule_partial} for multilayer-layer  Chern insulator slab: an out-of-plane electric field induces linearly increasing layer-projected out-of-plane orbital magnetization by redistributing electrons among the chiral edge states of different layers.
    (b) Honeycomb lattice for 2D Haldane model on which the hopping $t_1$ and $t_2 \exp(i \phi)$ are marked. We study two ways of stacking: $AA$-stacking and $AB$-stacking. Unless mentioned ad hoc, we always use $t_1=1$, $t_2=1/3$, $\phi=\pi/3$, $t_3=0.2$, $t_4=0.15$ and the staggered mass term is zero. (c) Experimental proposal to observe the quantized  OME response in a device formed by alternately stacking spacer and three-layer MnBi$_2$Te$_4$ (3L MBT) Chern layers, whose lattice structure is also given in the right panel, using SQUID-on-tip technique \cite{meltzer-SOT-prapp-2017}.}
    \label{fig:3dchern_schema}
\end{figure*}

Nevertheless, nothing forbids 3D Chern insulators to have a topological OME response. Intuitively, let us consider a 3D Chern insulator with Chern vector $\textbf{C} = (0,0,C_z)$, where $C_z = C (C\in \mathbb{Z}, C\neq 0)$ with interlayer distance $d$. On the one hand, it is straightforward to see that the quantized anomalous Hall current in response to weak in-plane electric field would induce  an off-diagonal OME response with its coefficient varying linearly in real space (see Sec.~\VII\ in Supplemental Materials \cite{supp}). On the other hand, if an out-of-plane uniform electric field ($\mathcal{E}_z$) is applied, the surface-state occupation would be different for every layer due to the vertical electrostatic potential drop. Then, the uneven distribution of electrons among the chiral edge states of each weakly coupled Chern-insulator layer would cause a layer-resolved orbital magnetization ($M_z$), which may manifest as a new type of diagonal OME effect unique to 3D Chern insulators, as schematically shown in Fig.~\ref{fig:3dchern_schema}(a). In this work, we find a never-mentioned quantization rule existing for the layer-resolved OME coupling of 3D Chern insulators: the difference between layer-resolved OME response coefficients is exactly quantized in units of $-C e^2/h$ (see Eq.~\eqref{alpha_zz_decoupled} and \eqref{eq:quantization_rule_partial}). Such quantization rule is exact in the presence of various types of interlayer couplings and stacking ways, and even remains robust against breaking of crystalline symmetries and disorder. Furthermore, through extensive first principles density functional theory (DFT) calculations, we suggest Mn(Bi/Sb)$_2$Te$_4$[(Bi/Sb)$_2$Te$_3$]$_n$ ($n=0, 1, 2, ...$) and Cr-doped (Bi/Sb)$_2$Te$_3$ as two classes of promising material platform (see Fig.~\ref{fig:3dchern_schema}(c) for schematic illustration) to realize such quantized magnetoelectric response. We further propose a specific experimental workflow using state-of-the-art magnetometer techniques to reveal the quantization rule in quasi-3D Chern insulator slabs.  %we do not only suggest a promising material platform exhibiting the quantization rule but also propose an experimental workflow to reveal it.
%Moreover, by adopting the ``gauge-discontinuity" formalism of Chern-Simons OME coupling \cite{liu-gaugeCS-prb-2015}, this quantization rule can be  derived in a mathematically rigorous way \cite{prepare}, and can be interpreted as a new type of ``anomalous" Chern-Simons  OME coupling unique to 3D Chern insulators.
%we find that such quantized OME response may 
%Then, the chiral edge states of each Chern-insulator layer (in the weakly coupled situation) would cause a layer-resolved $M_z$, which may be manifested as a new type of diagonal OME effect in 3D Chern insulators, as schematically shown in Fig.~\ref{fig:3dchern_schema}. 

\section{Layer-resolved orbital magnetization}

\subsection{Definitions of layer-resolved orbital magnetization and layer Chern number}
Although the total diagonal $\alpha_{zz}$, only relevant in 3D case, is vanishing in the presence of mirror or inversion symmetry, 
the layer-resolved $\alpha_{zz}^{(l)}$ (with layer index $l$), which measures the change of layer-resolved orbital magnetization $M_z^{(l)}$ under a weak $\mathcal{E}_z$, may be finite. We first show the quantization rule for $\alpha_{zz}^{(l)}$ using the definition $\partial M_z^{(l)}/\partial \mathcal{E}_z$. Since $\mathcal{E}_z$ would break the translational symmetry in the $z$ direction, it is natural to consider a slab geometry. Then, the layer-resolved magnetization of the $l$-th layer $M_z^{(l)}$ can be defined by summing over all the quantum degrees of freedom  except the layer index , which is a straightforward extension from the local definition of orbital magnetization \cite{resta-orbitalM-prb16,resta-localM-prl13}.

%The definition of layer-resolved orbital magnetization is inspired from the existing definition of local orbital magnetization given by \cite{resta-localM-prl13, resta-orbitalM-prb16}. 
Specifically, we use two expressions for layer-resolved orbital magnetization in a slab. More details on their derivation can be found in Sec.~\I\  of Supplemental Materials \cite{supp}. The first expression is derived from classical electrodynamics:
\begin{equation}
	\label{eq:M_def_brute}
	\begin{aligned}
		M_z = \frac{-e}{2A}\sum_{E_n<\mu} \bra{\phi_n}\vr\times\textbf{v}\ket{\phi_n}\bigg|_z \, ,
	\end{aligned}
\end{equation}
where  $\vr$ is the position, $\textbf{v}$ is the velocity, $A$ is the total area, $e>0$ is the elementary charge, and $\ket{\phi_n}$s are the eigenstates with energy $E_n$ below chemical potential $\mu$. Note that this expression only applies to systems with open boundary condition (OBC). This is because position operators are ill-defined in the thermodynamic limit and terms like $\bra{\phi_n}\vr\ket{\phi_n}$ diverge. This formula is particularly useful when we introduce disorder in the system in OBC. It contains both the contribution of circulation of the current density from the bulk and the contribution from the surface. Then, one can define layer-resolved orbital magnetization as 
\begin{equation}
    \label{eq:Mzl_def_brute}
    \begin{aligned}
        M_z^{(l)} = \frac{e}{\hbar A} {\rm Im} \sum_{\RR,s} \bra{s l,\RR} PxHyP \ket{s l,\RR},
    \end{aligned}
\end{equation}
where the atomic lattice basis $\ket{sl,\RR}$ represents the Wannier basis state at layer $l$, sublattice $s$ of unit-cell $\RR$. Here, $P$ is the projection operator onto the occupied subspace.

In the thermodynamic limit with periodic boundary conditions, we need to define a layer-projected orbital magnetization $M_z^{(l)}$ using the expression of total orbital magnetization \cite{resta-localM-prl13,resta-orbitalM-prb16}
\begin{equation}
\label{M_OBC_Resta}
    M_z = \frac{e}{\hbar A}{\rm Im\,Tr}\{P x Q(H-\mu)Q y P-Q x P(H-\mu)PyQ\},
\end{equation}
where $Q = 1 - P$ is the projection operator onto the unoccupied subspace. The chemical potential $\mu$ is at charge neutrality and $x, y$ are 2D position operators. ``Tr" in the above equation means taking the trace over the whole Hilbert space. Then, we can express $M_z$ in the Bloch eigen-function basis
\begin{equation}
\label{psi_layer_resolved}
    \ket{\psi_{n\textbf{k}}} = \sum_s \sum_{l=1}^{N_{\rm L}} C_{s l,n}(\textbf{k}) \ket{s l,\textbf{k}},\quad \textbf{k} = (k_x, k_y) \, ,
\end{equation}
where $s$ means sublattice index in the unit cell, $l$ layer index, $N_{\rm L}$ number of 2D layers in the $z$ direction, $n$ band index with energy $E_{n\textbf{k}}$ at wavevector $\textbf{k}$, and $\ket{s l ,\mathbf{k}}$ is the Fourier transform of the Wannier basis states $\ket{sl,\RR}$. In this basis, we can define a layer-projected orbital magnetization $M_z^{(l)}$
\begin{widetext}
\begin{equation}
\label{M_layer_resolved}
\begin{aligned}
    M_z^{(l)}=& \frac{e}{2\hbar A}{\rm Im}\big \{\epsilon_{z\alpha\beta}\sum_{s}\sum_\textbf{k}\sum_{\substack{n,n'\in{\rm o},\\ m\in{\rm e} }} C_{s l,n}(\textbf{k})C_{s l,n'}^*(\textbf{k})  \times(E_{m\textbf{k}}-\mu)\bra{u_{n\textbf{k}}}\partial_{k_\alpha}u_{m\textbf{k}}\rangle\bra{\partial_{k_\beta}u_{m\textbf{k}}}u_{n'\textbf{k}}\rangle\big \}\\
    &-\frac{e}{2\hbar A}{\rm Im}\big \{\epsilon_{z\alpha\beta}\sum_{s}\sum_\textbf{k}\sum_{\substack{m,m'\in{\rm e},\\ n\in{\rm o}} } C_{s l,m}(\textbf{k})C_{s l,m'}^*(\textbf{k})  \times(E_{n\textbf{k}}-\mu)\bra{u_{m\textbf{k}}}\partial_{k_\alpha}u_{n\textbf{k}}\rangle\bra{\partial_{k_\beta}u_{n\textbf{k}}}u_{m'\textbf{k}}\rangle\big \} \, ,
\end{aligned}
\end{equation}
\end{widetext}
where $A$ is the system area, $\rm o/e$ for occupied/empty states, $\ket{u_{n\textbf{k}}}$ is the periodic part of the Bloch eigenstate $\ket{\psi_{n\textbf{k}}}$ and  $\epsilon_{z\alpha\beta}$ with $\alpha,\beta=x,y$ is the antisymmetric tensor.

In the same spirit, we can also define layer Chern number in the Bloch basis
\begin{widetext}
\begin{equation}
    \label{eq:layer_chern}
\begin{aligned}
    C_z &= -\frac{1}{4 \pi}\epsilon_{z\alpha\beta}\,{\rm Im \, Tr}\{Pr_\alpha Qr_\beta P-Qr_\alpha Pr_\beta Q \}\\
    &=-\frac{\pi}{A}\epsilon_{z\alpha\beta}\,{\rm Im} \left\{\sum_{s\,l}\sum_{n,n'\in{\rm occ}}\sum_{m\in{\rm emp}}\sum_{\k }C_{sl,n}(\k )C_{sl,n'}^*(\k )\bra{u_{n\k }}\partial_{k_\alpha}u_{m\k }\rangle\bra{\partial_{k_\beta}u_{m\k }}u_{n'\k }\rangle \right.\\
    & \left. \quad \quad -\sum_{s\,l}\sum_{n\in{\rm occ}}\sum_{m,m'\in{\rm emp}}\sum_{\k }C_{sl,m}(\k )C_{sl,m'}^*(\k )\bra{u_{m\k }}\partial_{k_\alpha}u_{n\k }\rangle\bra{\partial_{k_\beta}u_{n\k }}u_{m'\k }\rangle \right\}\\
    &\equiv\sum_{l}C_z(l).
\end{aligned}
\end{equation}
\end{widetext}
Similar definitions are also used in Ref.~\cite{varnava-axion-prb-2018,gu-QAHEaxion-natcom-2021,li-highSaxion-natcom2024}.

\subsection{Response of layer-resolved orbital magnetization to vertical electric field}

For illustration, we model the 2D Chern-layer constituents of 3D Chern insulator by 2D Haldane model. Without loss of generality, the Chern number for each layer is set to be $C=1$ and interlayer coupling is weak such that the bulk gap remains finite. Details of parameters are shown in Fig.~\ref{fig:3dchern_schema}(b) and we consider two types of stacking, $AA$ and $AB$, as illustrated in Fig.~\ref{fig:3dchern_schema}(b). The $\textbf{k}$ mesh is set to $100 \times 100$. The chemical potential $\mu$ is determined by looking for charge neutrality point in the same system with open boundary conditions. 

%Then, $\alpha_{zz}^{(l)}$ can be obtained by numerically taking the ratio between change of $M_z^{(l)}$ and $\mathcal{E}_z$.  
In a slab system, we calculate $\alpha_{zz}^{(l)}$ using $\partial M_z^{(l)} / \partial \mathcal{E}_z$. We employ two approaches to calculate $\alpha_{zz}^{(l)}=\partial M_z^{(l)} / \partial \mathcal{E}_z$ in a slab system. The first one is to calculate $M_z^{(l)}$ in the absence and in the presence of weak electric field using Eq.~\eqref{M_layer_resolved}, then get $\alpha_{zz}^{(l)}$ by finite difference. In practice, we calculate $M_z^{(l)}$ for a series of weak electric field such that the response is still in the linear regime. Then, we fit the results to get the slope of the straight line and thus $\alpha_{zz}^{(l)}$. We use this approach for a slab using Eq.~\eqref{M_layer_resolved}, in which the wavefunctions and eigenergies are calculated using periodic boundary condition (PBC) in the $x,y$ plane. But, the chemical potential appearing in Eq.~\eqref{M_layer_resolved} is still determined using a slab with open boundary condition (OBC) in all three spatial directions with $100\times 100$ sites in the $(x,y)$ plane due to the presence of gapless surface states.

The second approach is to calculate $\alpha_{zz}^{(l)}$ directly applying perturbation theory to Eq.~\eqref{M_layer_resolved}. 
Specifically, when a vertical electric field is applied to a slab, our perturbed Hamiltonian reads: $H = H_0 +e \mathcal{E}_z z$. Suppose the electric field is so weak such that bulk gap is not closed and the potential drop between the top and bottom layer is the smallest energy scale in the system. To get a perturbation theory of layer-resolved OME response, we need to express Eq.~\eqref{M_layer_resolved} of perturbed Hamiltonian $H$ using the information of $H_0$. By singling out the terms proportional to $\mathcal{E}_z$, the layer-resolved OME response $\alpha_{zz}^{(l)}$ is then derived by $\partial M_z^{(l)} / \partial \mathcal{E}_z$. After gathering all the first-order corrections (including energy, wavefunction and chemical potential with particle number fixed), a semi-analytic formula for $\alpha_{zz}^{(l)}$ can be derived. Details of the derivation are given in Sec.~\III\  of Supplemental Materials \cite{supp}. We say ``semi-analytic'' since the determination of the chemical potential still requires solving numerically the system with open boundary condition, which is unavoidable due to the presence of gapless surface states in 3D Chern insulators. 

Compared to the first approach, one advantage of the perturbative approach is that the quantization gradually improves as the $\mathbf{k}$ mesh becomes finer, as will be presented in the following. Another advantage of the perturbation approach is that the mirror symmetry is enforced in the results: $\alpha_{zz}^{(l)}$ is rigorously symmetric with respect to the central plane.

%The third approach is similar to the first approach except that we calculate $M_z^{(l)}$ using Eq.~\eqref{eq:Mzl_def_brute}. This method is uniquely suitable for a slab with OBC in the $x,y$ plane. We use this method to test the robustness of the quantization rule against disorder. 

We apply the above two numerical approaches to multilayer Haldane model with both $AA$ and $AB$ stackings. It turns out that $M_z^{(l)}$ varies linearly for weak electric field for both types of stacking (see Fig.~S1 in Supplemental Materials \cite{supp}). The obtained $\alpha_{zz}^{(l)}$ is quantized to integers if the number of layers $N_{\rm L}$ is odd or to half-integers if $N_{\rm L}$ is even, as shown in Table~\ref{Table_layer_resolved_PBC}. The distribution is symmetric to the central plane of the system and the total $\alpha_{zz}$ is indeed vanishing as dictated by mirror symmetry. The quantization is limited to $1\%$ precision due to the finite-size effect \cite{supp}. If $\mathcal{E}_z$ is so weak that the electrostatic potential drop between the top and bottom layers can be treated as a perturbation to the system, then $\alpha_{zz}^{(l)}$ can be determined semi-analytically using perturbation theory as described above. Therefore, one can directly calculate $\alpha_{zz}^{(l)}$ using only the energy and Bloch functions of the unperturbed system. As shown in the parenthesis of Table~\ref{Table_layer_resolved_PBC}, $\alpha_{zz}^{(l)}$ obtained by semi-analytic perturbation theory follow exactly the same quantization rule as those calculated by numerical finite-difference method. 

%Since the layer index $l$, the choice of which is a priori arbitrary, appears in perturbation formula \cite{supp}, the final formula seems to depend on the ``layer index gauge choice'' at the first glance. Eventually, we can prove that the layer-resolved responses turns out to be ``layer index gauge independent'' as proved in Sec.~\IV of Supplemental Materials \cite{supp}. 

\begin{table*}%[H] add [H] placement to break table across pages
\caption{Layer-resolved $\alpha_{zz}^{(l)}$ in units of $e^2/h$ obtained by numerical finite difference and those evaluated by perturbation theory given in the parenthesis. We consider $AA$,$AB$ stackings and $N_{\rm{L}} = 5,6$.}
\label{Table_layer_resolved_PBC}
\begin{ruledtabular}
\begin{tabular}{c| c c c c c c}
Layer Index &1&2&3&4&5&6\\
\hline
AA,5&1.9893 (1.9905)&0.9971 (0.9982)&-0.0010 (9.80E-14)&-0.9990 (-0.9982)&-1.9918 (-1.9905)&\diagbox[dir=SW]{}{}\\
AB,5&1.9289 (1.9739)&0.9525 (0.9973)&-0.0456 (5.29E-05)&-1.0424 (-0.9974)&-2.0194 (-1.9739)&\diagbox[dir=SW]{}{}\\
AA,6&2.5365 (2.4894)&1.5444 (1.4974)&0.5464 (0.4992)&-0.4519 (-0.4992)&-1.4502 (-1.4974)&-2.4423 (-2.4894)\\
AB,6&2.4863 (2.4727)&1.5098 (1.4962)&0.5123 (0.4987)&-0.4855 (-0.4990)&-1.4828 (-1.4963)&-2.4592 (-2.4727)\\
\end{tabular}
\end{ruledtabular}
\end{table*}

\begin{table*}%[H] add [H] placement to break table across pages
    \caption{Difference between $\alpha_{zz}^{(l)}$ of neighboring layers in units of $e^2/h$ obtained by numerical finite difference and those evaluated by perturbation theory given in the parenthesis. We consider $AA$,$AB$ stackings and $N_{\rm{L}} = 5,6$.}
    \label{Table_diff_PBC}
    \begin{ruledtabular}
    \begin{tabular}{c| c c c c c c}
    $\alpha_{zz}^{(l+1)}-\alpha_{zz}^{(l)}$ &$l=1$&$l=2$&$l=3$&$l=4$&$l=5$\\
    \hline
    AA,5&-0.9921(-0.9922)&-0.9982(-0.9982)&-0.9980(-0.9982)&-0.9927(-0.9922)&\diagbox[dir=SW]{}{}\\
    AB,5&-0.9794(-0.9766)&-0.9981(-0.9972)&-0.9968(-0.9972)&-0.9770(-0.9766)&\diagbox[dir=SW]{}{}\\
    AA,6&-0.9921(-0.9920)&-0.9980(-0.9982)&-0.9983(-0.9983)&-0.9983(-0.9982)&-0.9921(-0.9920)\\
    AB,6&-0.9764(-0.9766)&-0.9976(-0.9975)&-0.9978(-0.9977)&-0.9973(-0.9972)&-0.9764(-0.9765)\\
    \end{tabular}
    \end{ruledtabular}
\end{table*}
% \paragraph{A limiting case that interlayer decoupled --} 
To understand the origin of the quantization in $\alpha_{zz}^{(l)}$, one can consider the simplest case where neighboring layers are completely decoupled from each other such that the Hamiltonian is block diagonal in layer space. Then, we can prove rigorously that the difference in $\alpha_{zz}^{(l)}$ between the two neighboring layers is exactly quantized (see Sec.~\V\  of Supplemental Materials \cite{supp})
\begin{equation}
    \label{alpha_zz_decoupled}
        \alpha_{zz}^{(l+1)} - \alpha_{zz}^{(l)} = -C\,\frac{e^2}{h}\, ,
\end{equation}
or equivalently,
\begin{equation}
\frac{\partial\alpha_{zz}(l)}{\partial l}= -C\,\frac{e^2}{h} \, .
\label{eq:quantization_rule_partial}
\end{equation}
This quantization rule remains robust even in the presence of considerable interlayer hopping, as shown in Table~\ref{Table_diff_PBC} with $C=1$. Knowing that the total $\alpha_{zz}$ is vanishing, we can deduce that $\alpha_{zz}^{(l)}$ is integer quantized in units of $e^2/h$ for odd number of layers or half-integer quantized for even number of layers, consistent with Table~\ref{Table_layer_resolved_PBC}. 

The quantized $\partial_l \alpha_{zz}(l)$ in 3D Chern insulator is attributed to topological edge states. Intuitively, 3D Chern insulator is formed here by stacking $N_{\rm L}$ layers of $C=1$ 2D Chern insulators. There must be $N_{\rm L}$ gapless chiral edge states and a charge neutral $\mu$ crosses with all the edge states. Applying weak uniform $\mathcal{E}_z$ induces different occupations for the edge states of different layers when the system is back to equilibrium. The redistribution of electrons among different layers can be equivalently modelled by layer-resolved chemical potential $\mu^{(l)} = \mu_{0}-e\mathcal{E}_z d (l-\Delta l_{N_{\rm L}})$, where $\mu_{0}$ is the chemical potential in the absence of electric field and $\Delta l_{N_{\rm L}}=(N_{\rm L}+1)/2$ such that the total number of electrons is unchanged, namely $\sum \mu^{(l)} =N_{\rm L}\mu_0$. Since the chemical potential is in the bulk gap, all the changes in $M_z^{(l)}$ stems from the surface-state contribution \cite{vanderbilt-berry-book-2018, resta-orbitalM-prb16,resta-orbitalM-prl05, resta-orbitalM-prb06,seleznez-surfaceM-prb-2023}:
\begin{equation}
    M_z^{(l)} = \mu^{(l)} \frac{e}{h d} C_z (l) \, ,
\end{equation}
where the division by $d$ makes $M_z^{(l)}$ volumetric. Therefore, we can get a simple formula for $\alpha_{zz}^{(l)}$ 
\begin{equation}
    \alpha_{zz}^{(l)} = \frac{\partial M_z^{(l)}}{\partial\mu^{(l)}}\frac{\partial\mu^{(l)}}{\partial\mathcal{E}_z} =-C_z (l) \frac{e^2}{h} (l-\Delta l_{N_{\rm L}}),
\end{equation}
consistent with Eq.~\eqref{alpha_zz_decoupled} given that $C_z (l)=C$. The behavior for even and odd number of layers are precisely given by the shift $\Delta l_{N_{\rm L}}$ as dictated by mirror symmetry. This picture is exact in the interlayer decoupled case and generalizable to the interlayer coupled case by the argument that as long as the bulk gap is not closed, the layer Chern number (see Methods for definition) remains quantized $C_z (l)=C$ \cite{gu-QAHEaxion-natcom-2021,rauch-sAHE-prb-2018,varnava-axion-prb-2018,resta-localchern-prb11,li-highSaxion-natcom2024}, so that the quantized OME response protected by the layer Chern number should remain robust.

\section{Layer-resolved electric polarization}
The same quantization rule also exists for a genuine 3D Chern insulator in which the vertical stacking of 2D Chern layers goes to infinity with periodic boundary condition. To evaluate $\alpha_{zz}^{(l)}$, it is convenient to calculate the change of layer-resolved polarization $P_{z}^{(l)}$ \cite{wu-layerP-prl-2006} under a finite weak magnetic field $B_z$. By adopting a magnetic field commensurate with crystalline lattice, we restore the translational symmetry in all the three directions such that $\alpha_{zz}^{(l)}$ is determined in the thermodynamic limit without explicitly invoking finite-size effect in the slab geometry. %However, as shown below, the quantization of $\alpha_{zz}^{(l)}$ still remain robust, which  is deeply related to chiral edge states via Streda formula \cite{streda-1982}. 

Before calculating $\alpha_{zz}^{(l)}$, we need to properly define the electronic contribution of layer polarization $P_{z}^{(l)}$ under weak vertical magnetic field $B_z$, which can be written in terms of the maximally localized Wannier center for each layer in the stacking direction, denoted as $\bar{z}_l$ \cite{wu-layerP-prl-2006}
%The total polarization can be decomposed into ionic and electronic contribution. We only focus on the latter so that 
\begin{equation}
    P_{z}^{(l)} = - \frac{e}{\Omega d} \bar{z}_l
    \label{Eq: Pzl}
\end{equation}
where $\Omega$ is the area of the 2D magnetic primitive cell, interlayer distance $d$ is constant, and the averaged maximally localized Wannier center is proportional to the averaged Berry phase $\phi_l$ calculated by parallel-transport method \cite{wu-layerP-prl-2006,marzari-wannier-prb-1997}
\begin{equation}
    \bar{z}_l = \frac{N_{\text{L}} d}{N_k} \sum_{k} \frac{-\phi_l}{2 \pi} \,.
    \label{Eq: Wannier_zl}
\end{equation}
where $N_k$ is the total number of $\textbf{k}$ points in the $x$-$y$ plane. In practice, parallel-transport method gives $N_{\text{L}} N_k$ Berry phases, which are then categorized into $\phi_l$ of different layers based on the smallest distance criterion \cite{wu-layerP-prl-2006}. Here $N_{\rm{L}}$ is the number of layers in an artificially chosen repetition slab, namely a supercell in the $z$-direction. We consider infinite stacking of these slabs (formed by 2D Haldane model as before) with periodic boundary condition imposed, so that the vertical size of 3D Brillouin zone is set to $2\pi/(N_{\rm{L}}d)$ with $d$ being interlayer distance.

It turns out that even with a relative strong interlayer coupling (interlayer hopping $t_3,t_4$ is of the same order magnitude as in-plane hopping $t_2$), the Wannier centers still sit closely to each layer (see Sec.~\VI\  of Supplemental Materials \cite{supp}). In practice, we use a magnetic field such that the magnetic flux piercing the atomic unit cell is $p/q=1/30$. A $48 \times 48 \times 48$ \textbf{k} mesh is adopted for numerical calculations of the Hofstadter states under magnetic field and the corresponding Berry phases. 
For the sake of consistency, we also use the same magnetic unit cell in the absence of magnetic field. Then, we can calculate the layer-resolved polarization using Eq.~\eqref{Eq: Pzl}. After doing a finite difference between $P_z^{(l)}$ in the absence and in presence of magnetic field, we average over the 2D magnetic Brillouin zone to get $\alpha_{zz}^{(l)}$.
In the end, we obtain a similar quantization rule for $\alpha_{zz}^{(l)}$ %in the thermodynamic limit 
as shown in Table~\ref{Table_bulk} given $C=1$. Comparing with previous calculations, the quantization is much better ($\sim 0.1 \%$) in such situation with fully periodic boundary condition, even in the presence of relatively strong interlayer couplings. 

\begin{table*}[!htbp]
    \caption{Layer-resolved response in units of [$e^2/h$] in the thermodynamic limit using $\partial P/ \partial B$.}
    \label{Table_bulk}
    \begin{ruledtabular}
    \begin{tabular}{c| c c c c c}
    Layer Index &1&2&3&4&5\\
    \hline
    AA,4&1.499985&0.50002&-0.49998&-1.50002&\diagbox[dir=SW]{}{}\\
    AA,5&1.999985&1.00002&0.00002&-1.00000&-2.00001\\
    AB,4&1.50401&0.49599&-0.50555&-1.49445&\diagbox[dir=SW]{}{}\\
    \end{tabular}
    \end{ruledtabular}
\end{table*}

%By Streda formula \cite{streda-1982}, when a finite magnetic field is turned on, states would flow between valence and conduction bands while keeping the chemical potential in the bulk gap. The number of transferred states is proportional to the Chern number of the system. For example, for $N_{\text{L}}$ = 4, when $B_z$ is switched on, four states would flow from conduction to valence bands as dictated by four $C=-1$ 2D Chern layers. Moreover, %Most saliently, we observe that 
%these four states are tied exactly to four equidistant Wannier centers from the four coupled 2D Chern layers, respectively. 
%Indeed, the quantization rule remains robust as long as the extra $B_z$-induced flowing states are evenly distributed among the $N_{\rm{L}}$ equidistant Wannier centers, which do not necessarily coincide with the layer position. The validity of Eq.~\eqref{alpha_zz_decoupled}  even do not require uniform layer charge density when $B_z=0$. The only requirement is that the bulk gap remains opened and layer Chern number remains exactly quantized as interlayer coupling is turned on, which is obvious for disorder-free 3D Chern insulators where $k_z$ is a good quantum number.

%Following the above argument, 
Here we prove that the quantization rule Eq.~\eqref{alpha_zz_decoupled} for interlayer-coupled 3D Chern insulators relies on the quantization of layer Chern number. First, due to the Streda formula \cite{streda-1982}, under finite $B_z$, the layer charge density $n_l$ follows $\partial n_l/\partial B_z=-C_z(l) e^2/h$ for all layers, with $C_z(l)$ denoting layer Chern number. Since layer Chern numbers are all equal to each other for a 3D Chern insulator, whenever there is a change in the total charge density (of the $N_{\rm{L}}$-layer unit cell) induced by $B_z$, it has to be evenly distributed to the $N_{\rm{L}}$ equidistant averaged Wannier centers, as long as $C_z(l)$ is still well defined and quantized. The additional layer charge density leads to a change in layer-resolved polarization $\delta P_z^{(l)}=\delta n_l  \overline{z}_l/d $ according to Eq.~\eqref{Eq: Pzl}, which is induced by $B_z$. Thus, the layer-resolved OME coupling is given by:
\begin{equation}
    \alpha_{zz}^{(l)} = \frac{\delta P_{z}^{(l)}}{\delta n_l} \frac{\delta n_l}{\delta\!B_z} = -\frac{\overline{z}_l}{d} C_z(l) \frac{e^2}{h}\,.
\end{equation} 
Note that the average Wannier center $\overline{z}_l$ does not necessarily sit at the corresponding layer position due to lack of crystalline symmetry. However, by translation symmetry in the $z$-direction, we must have $\overline{z}_l= l d +\Delta z$ with constant interlayer distance $d$. Then, Eq.~\eqref{alpha_zz_decoupled} follows immediately. Here, it may be more suitable to say ``translation symmetry of interlayer coupling'', namely two neighboring layers are always coupled in the same way, because $AB$ stacking does not impose one-layer but two-layer translational symmetry. Nevertheless, the calculations of the Wannier center average over two sublattices so that two stackings make no difference on $\overline{z}_l$. So, $\overline{z}_l$ must be equidistantly distributed. The importance of translation symmetry can also be seen from Table~\ref{Table_diff_PBC} where the quantization for two outmost layers is worse than the inner layers for $AB$ stacking. Furthermore, the validity of Eq.~\eqref{alpha_zz_decoupled} even does not require uniform layer charge density when $B_z=0$. The only requirement is that the bulk gap remains opened and layer Chern number remains exactly quantized as interlayer coupling is turned on, which is obvious for disorder-free 3D Chern insulators where $k_z$ is a good quantum number.

\section{Robustness of the quantization rule}

\subsection{Robustness against symmetry breaking and substrate effects}
To illustrate the case without inversion and mirror symmetries, let us go back and consider a slab and introduce a different on-site energy only to the first layer (denoted as ``substrate" layer) such that the total response $\alpha_{zz}$ becomes finite due to the breaking of mirror symmetry. 
For illustration, we consider a slab of five $AA$-stacked layers. We set the on-site energy of the first layer to be $1/6$ while varying the amplitude of $t_{3,4}$. We calculate $\alpha_{zz}^{(l)}$ using Eq.~\eqref{M_layer_resolved} with $90 \times 90$ $\mathbf{k}$ mesh. Although $\alpha_{zz}^{(l)}$ itself does not follow any quantization rule (see Table~S1 in Supplemental Materials \cite{supp}), the difference between the neighboring layer-resolved responses is quantized to layer Chern number as predicted, as shown in Table~\ref{Table_substrate}. This shows that the most intrinsic quantization rule is Eq.~\eqref{alpha_zz_decoupled}. Nevertheless, symmetry can impose additional constraints on the values of $\Delta z$. For example, mirror or inversion symmetry forces $\Delta z$ to be integer or half integer, which thus enforces $\alpha_{zz}^{(l)}$ to be integer or half integer values, as shown in Table~\ref{Table_layer_resolved_PBC} and \ref{Table_bulk}.

\begin{table*}[!htbp]
    \centering
    \caption{Difference between neighboring layer-resolved response in units of [$e^2/h$] in systems with substrate.}
    \label{Table_substrate}
    \begin{ruledtabular}
    \begin{tabular}{c| c c c c}
    $\alpha_{zz}^{(l+1)}-\alpha_{zz}^{(l)}$ & $l=1$ & $l=2$ & $l=3$ & $l=4$ \\
    \hline
    $t_3=0,t_4=0$ & -0.9972&-0.9972&-0.9972&-0.9972\\
    % \cline{2-2}\cline{5-9}
    $t_3=1/6,t_4=0$ & -0.9972&-0.9972&-0.9972&-0.9972\\
    % \cline{2-2}\cline{5-9}
    $t_3=0.3,t_4=0$ & -0.9972&-0.9972&-0.9972&-0.9972\\
    % \cline{2-2}\cline{4-9}
    $t_3=1/6,t_4=3t_3/4$ & -0.9934&-0.9978&-0.9978&-0.9937\\
    % \cline{2-2}\cline{5-9}
    $t_3=0.3,t_4=3t_3/4$ & -0.9840&-0.9996&-0.9990&-0.9849\\
    \end{tabular}
    \end{ruledtabular}
\end{table*}

Therefore, we see from above that the quantization rule Eq.~\eqref{alpha_zz_decoupled} is precisely given by the integer number of transferred states projected to each layer induced by magnetic field, which essentially originate from the quantized layer Chern number and topological edge states. This reveals again the topological nature of the quantization in $\alpha_{zz}^{(l)}$.

\subsection{Robustness against disorder}
\begin{table*}[!htbp]  %[H] add [H] placement to break table across pages
	\caption{Statistical layer-resolved response in units of [$e^2/h$] in the presence of disorder using $\partial M/ \partial \mathcal{E}$ (fluctuation $\equiv$ std($15-1$)).}
	\label{Table_disorder}
	\begin{ruledtabular}
	\begin{tabular}{c| c c c c c c}
	layer index&2&1&0&-1&-2\\
	\hline
	bare&1.9401&0.9934&-0.0034&-0.9875&-1.9426\\
	Type-I average&1.9398&0.9932&-0.0034&-0.9874&-1.9422\\
	Type-II average&1.8965&1.0213&0.0512&-0.9527&-2.0127\\
	Type-I fluctuation&2.73$\times 10^{-4}$&1.47$\times 10^{-4}$&1.34$\times 10^{-5}$&1.37$\times 10^{-4}$&2.82$\times 10^{-4}$\\
	Type-II fluctuation&0.0443&0.0584&0.0600&0.0286&0.0407\\
	\end{tabular}
	\end{ruledtabular}
\end{table*}

In the presence of disorder, as long as the bulk gap is not closed by disorder such that layer Chern number remains well defined and quantized, the change of layer charge density induced by magnetic field would still remain quantized as dictated by the Streda formula. Thus, the quantization rule of Eq.~\eqref{alpha_zz_decoupled} is still valid. More numerical evidence of the robustness of Eq.~\eqref{alpha_zz_decoupled} in the presence of disorder is given in Table ~\ref{Table_disorder}. 
In practice, we need to calculate $\alpha_{zz}^{(l)}$ numerically by finite difference in the  open boundary conditions. Disorder is modelled by random on-site potential. Two types of disorder are introduced. The first type is that on-site energy only fluctuates in-plane repeating itself among all the layers. The second type breaks the translation symmetry such that on-site energy fluctuates among all the sites of all layers. We consider 15 disorder configurations and then analyze the results statistically. Here we set $t_3 = 0.2$, $t_4 = 0$, and consider a strong disorder with amplitude $0.4t_3$. As shown in Table~\ref{Table_disorder}, the quantization rule remains good statistically not only for Type-\I\ disorder preserving layer translation symmetry but also Type-\II\ disorder breaking it. This is because the amplitude of the disorder does not close the gap so that layer Chern number can still be well defined. 

\section{Experimental realizations}

The considerable and quantized difference between layer-resolved OME $\alpha_{zz}^{(l)}$ is  measurable in a slab system. As schematically shown in Fig.~\ref{fig:3dchern_schema}(c), a slab system can be formed by stacking vertically 2D Chern insulators. The interlayer coupling  is in principle tunable, as already experimentally achieved by inserting a spacer with varying thickness between two adjacent Chern layers \cite{zhao-highCQAHE-nature-2020}. Then, the quantized OME response can be experimentally measured in such Chern-insulator slab using state-of-the-art magnetometer techniques such as single-spin scanning magnetometry using nitrogen vacancy in diamond \cite{thiel-NVcenter-science-2019,tschudin-NVcenterCrSBr-natcom-2024} and SQUID-on-tip sensors \cite{uri-QHEedgeSOT-natphys-2020,zhou-squidGr-nature-2023}, where ``SQUID” is an abbreviation of Superconducting Quantum Interference Device. 

%\sout{between which a topologically trivial insulating spacer needs to be inserted to avoid bulk gap closure, which would be induced by strong interlayer coupling.}

\subsection{Material proposals}
Before providing a specific experimental approach to reveal the topological OME response presented above, we first suggest two possible material systems which can form a slab of quasi-3D Chern insulator. One promising material platform is $\rm{(Bi/Sb)}_2 \rm{Te}_3$ thin film, which is 2D Chern insulator if stacking three Cr-doped quintuple layers \cite{zhao-highCQAHE-nature-2020}. It has been shown in the experiments \cite{zhao-highCQAHE-nature-2020} that quasi-3D Chern insulator slab can be synthesized by alternately growing three Cr-doped quintuple layers and four undoped ones (served as spacer) in the $z$-direction, which exhibit well quantized anomalous Hall resistance with high Chern numbers thanks to good quality of samples. 

\begin{figure*}
    \centering
    \includegraphics[width=0.8\textwidth]{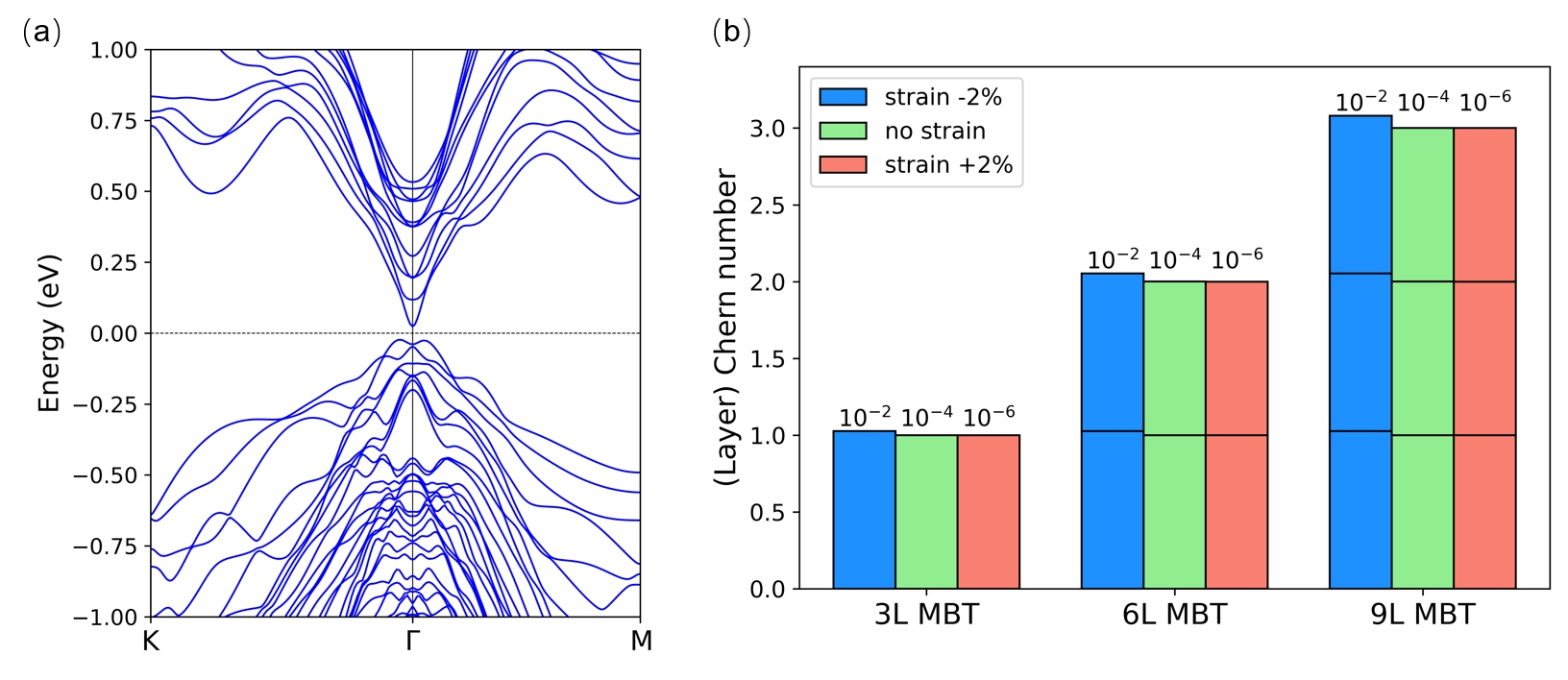}
    \caption{(a) Band structure of 3L MBT without strain in which the zero energy indicates chemical potential and the indirect gap is 46 meV. (b) Layer Chern number and total Chern number for 3,6,9-layer MBT in the presence of in-plane strain -2\%, 0\% and 2\%. The height of each sub-bar indicates the calculated layer Chern number for each 3L MBT entity using Eq.~\eqref{eq:layer_chern} under the same $150 \times 150$ $\mathbf{k}$-mesh. The total height is then the total Chern number. The error of the quantization is given above each bar. We see that the quantization of the calculated Chern numbers is better with strain 2\% but worse with strain -2\% because the indirect gap increases with strain from -2\% to 2\% (see Sec.~\VIII\  of Supplemental Materials \cite{supp}).}
    \label{fig:MBT}
\end{figure*}

Another candidate is the Mn(Bi/Sb)$_2$Te$_4$[(Bi/Sb)$_2$Te$_3$]$_n$-class ($n=0, 1, 2, ...$) of materials. To get a building block of 3D Chern insulator, we explore several kinds of Mn(Bi/Sb)$_2$Te$_4$[(Bi/Sb)$_2$Te$_3$]$_n$-type thin films with different number of layers and different in-plane strain magnitudes using first principles DFT calculations (see Sec.~\VIII\ of Supplemental Materials). As shown in Table~\ref{Table_MBT}, the most stable case is  3-layer $\rm{MnBi}_2 \rm{Te}_4$ in ferromagnetic state, which is a $C=1$ Chern insulator and remains topologically nontrivial over a large range of in-plane strain amplitudes from -2\% to 2\%. The indirect gap of such Chern insulator is as large as $46\,$meV as shown in Fig.~\ref{fig:MBT}(a). %Such large gap guarantees the non-trivial topology against strain and other disorders, namely the Chern number of three-layer $\rm{MnBi}_2 \rm{Te}_4$ would not easily change, making it a promising building block for 3D Chern insulator slab. 
Such a large topological gap makes 3-layer $\rm{MnBi}_2 \rm{Te}_4$ a promising building block for quasi-3D Chern-insulator slab.
For illustration, we construct the Wannier tight-binding Hamiltonian (based on DFT calculations) for 3, 6, and 9-layer $\rm{MnBi}_2 \rm{Te}_4$ slabs assuming a proper spacer is inserted between every two adjacent three-layer $C=1$ entities, so that the inter-entity coupling can be determined from DFT calculations (see Sec.~\VIII\ of Supplemental Materials). %As shown in Fig.~\ref{fig:device_chern}(b), 
As expected, the calculated total Chern numbers for the 3, 6, 9-layer $\rm{MnBi}_2 \rm{Te}_4$ slabs are $C=1, 2, 3$, respectively. The calculated layer Chern numbers are always well quantized to 1 as shown in Fig.~\ref{fig:MBT}(b), even under substantial in-plane strain. This suggests that MnBi$_2$Te$_4$ thin films with properly intercalated interlayer spacer is an ideal platform to test the quantization rule Eq.~\eqref{alpha_zz_decoupled}.  %Usually, inserting insulator spacer (such as hexagonal boron nitride) would unavoidably introduce in-plane strain to the system. The robustness of the Chern gap against strain for 3-layer $\rm{MnBi}_2 \rm{Te}_4$ %So, even if we use hBN as spacer, which would create strain between spacer and Chern layer, layer Chern numbers of 3D Chern insulator slab is still well quantized and thus so does the quantization rule Eq.~\eqref{alpha_zz_decoupled}. 

\begin{table*}[!htbp]
    \centering
    \caption{Chern numbers of Mn(Bi/Sb)$_2$Te$_4$[(Bi/Sb)$_2$Te$_3$]$_n$-class thin films under different strain magnitudes. ``FM“ and ``AFM" stand for interlayer ferromagnetic and anti-ferromagnetic configurations, respectively.}
    \label{Table_MBT}
    \tabcolsep=3mm
    \renewcommand\arraystretch{1.5}
    \begin{tabular}{l|ccccccccc}
        \hline
        \hline
        $\mathrm{system\,(layers\mid magnetism\mid strain\,axis)}$& \multicolumn{1}{c|}{-2.0\%} & \multicolumn{1}{c|}{-1.5\%} & \multicolumn{1}{c|}{-1.0\%} & \multicolumn{1}{c|}{-0.5\%} & \multicolumn{1}{c|}{0.0\%} & \multicolumn{1}{c|}{0.5\%} & \multicolumn{1}{c|}{1.0\%} & \multicolumn{1}{c|}{1.5\%} & 2.0\% \\ \hline
        $\mathrm{\quad\quad\quad MnBi_2Te_4\,\,\,(\,2\mid \,\,\,FM\,\,\mid a\,)}$ & \multicolumn{8}{c|}{0}& 1\\ \hline
        $\mathrm{\quad\quad\quad MnBi_2Te_4\,\,\,(\,3\mid AFM\,\mid a\,)}$ & \multicolumn{8}{c|}{0}& 1\\ \hline
        $\mathrm{\quad\quad\quad MnSb_4Te_7\,\,\,(\,2\mid \,\,\,FM\,\,\mid a\,)}$ & \multicolumn{1}{c|}{1}& \multicolumn{8}{c}{0}\\ \hline
        $\mathrm{\quad\quad\quad MnSb_4Te_7\,\,\,(\,2\mid \,\,\,FM\,\,\mid c\,)}$ & \multicolumn{4}{c|}{1}& \multicolumn{5}{c}{0}\\ \hline
        $\mathrm{\quad\quad\quad MnSb_6Te_{10}\,(\,2\mid \,\,\,FM\,\,\mid a\,)}$ & \multicolumn{4}{c|}{0}& \multicolumn{5}{c}{1}\\ \hline
        $\mathrm{\quad\quad\quad MnBi_2Te_4\,\,\,\,(\,3\mid \,\,\,FM\,\,\mid a\,)}$  & \multicolumn{9}{c}{1} \\ \hline
        \hline
    \end{tabular}
\end{table*}

\subsection{Workflow for experimental detection}
Once the encapsulated device of quasi-3D Chern insulator slab is synthesized, one can use a magnetometer with high precision to acquire profiles of magnetic field $\mathbf{B}$ outside the device. The sensitivity of the state-of-art technique such as single-spin scanning magnetometry using nitrogen vacancy in diamond \cite{thiel-NVcenter-science-2019,tschudin-NVcenterCrSBr-natcom-2024} and SQUID-on-tip sensors \cite{uri-QHEedgeSOT-natphys-2020,zhou-squidGr-nature-2023} is below nano Tesla \cite{zhou-squidGr-rsins-2023}, through which topological magnetoelectric effect in the quantum Hall state has already been observed \cite{uri-QHEedgeSOT-natphys-2020}. Then, layer-resolved magnetization $M_z^{(l)}$ can be reconstructed by reverse propagation \cite{roth-currentReconstruct-japhys-1989,feldmann-currentReconstruct-prb-2004,meltzer-SOT-prapp-2017,broadway-magnReconstruct-prapp-2020}. We briefly discuss how this method can be implemented for our problem. 

We suppose that the measured magnetic field is solely generated by the magnetization of the device. The law of Biot-Savart in magnetostatics stipulates that
\begin{equation}
    \label{eq:biot-savart}
 \mathbf{B} 
 = \frac{\mu_0}{4\pi} \iiint d^3 r' \, \frac{ \mathbf{J}(\mathbf{r}') \times (\mathbf{r}-\mathbf{r}')}{|\mathbf{r}-\mathbf{r}'|^3}
\end{equation}
with $\mu_0$ is the permeability of vacuum and the current density $\mathbf{J}(\mathbf{r}')$ is bound current generated by magnetization $M$, $\mathbf{J} = \nabla \times \mathbf{M}$. Here, we neglect in-plane magnetization since Chern layers are atomically thin, and out-of-plane magnetization contributed by each Chern layer depends only on spatial coordinates $x$ and $y$, $\mathbf{M}= M (x,y) \hat{z}$. Let us take SQUID-on-tip technique as example, which probes only the $z$ component of magnetic field.  Direct calculations from Eq.~\eqref{eq:biot-savart} lead to a formula for magnetic field outside the device created by a Chern layer confined between $z_1$ and $z_2$
\begin{align}
    \label{eq:biot-savart-M}
    &B_z(x,y,z) \;\nn
    =& \iint d x' d y'\, K(x-x',y-y',z)\, M(x',y')
\end{align}
with the kernel function $K(x,y,z)$
\begin{align}
    \label{eq:kernel-r}
    &K(x,y,z) \;\nn
    =& \frac{\mu_0}{4\pi} \left( \frac{z-z_2}{(x^2+y^2+(z-z_2)^2)^{\frac{3}{2}}} - \frac{z-z_1}{(x^2+y^2+(z-z_1)^2)^{\frac{3}{2}}} \right) .
\end{align}
Eq.~\eqref{eq:biot-savart-M} can be nicely dealt under Fourier transformation such that convolution becomes trivial product
\begin{equation}
    \label{eq:biot-savart-M-FT}
    \tilde{B}_z(k_x,k_y,z) = \tilde{K}(k_x,k_y,z)\, \tilde{M}(k_x,k_y)
\end{equation}
where the notation with tilde is the Fourier transform of the corresponding function. It turns out that the kernel function $\tilde{K}(k_x,k_y,z)$ can be computed analytically
\begin{equation}
    \label{eq:kernel-k}
    \tilde{K}(k_x,k_y,z) = \frac{\mu_0}{2} e^{-k z}\left( e^{k z_2} - e^{k z_1}  \right)
\end{equation}
with $k=\sqrt{k_x^2+k_y^2}$. Therefore, as long as the profile of magnetic field at given $z$ is known, we can calculate its Fourier transform and then invert the kernel function in Eq.~\eqref{eq:biot-savart-M-FT} to get the profile of the magnetization in the Chern layer after inverse Fourier transformation. This process is called reverse propagation. 

Nevertheless, the measured magnetic field $B_z$ is generated by all the Chern layers in quasi-3D Chern insulator slab. Each Chern layer $l$ is equipped with its own kernel function $\tilde{K}_l$ where $z_{1,2}$ are replaced by the layer-resolved $z_{1,2}^{(l)}$ in Eq.~\eqref{eq:kernel-k}. So, Eq.~\eqref{eq:biot-savart-M-FT} becomes
\begin{equation}
    \label{eq:biot-savart-M-FT-layer}
    \tilde{B}_z(k_x,k_y,z) = \sum_{l=1}^{N_{\text{L}}} \tilde{K}_l (k_x,k_y,z)\, \tilde{M}_z^{(l)}(k_x,k_y)
\end{equation}
where $N_{\text{L}}$ is the total number of Chern layers and $\tilde{M}_z^{(l)}$ is the Fourier transform of  magnetization of layer $l$. To get $\tilde{M}_z^{(l)}$, we need to collect the data of $B_z$ at $N_{\text{L}}$ different height $h_i$ with $i=1,\dots,N_{\text{L}}$ outside the device. Then, reverse propagation becomes matrix inversion
\begin{align}
    \label{eq:biot-savart-M-FT-layer-matrix}
%    \tilde{B}_z(k_x,k_y,h_i) &= \sum_{l=1}^{N_l} \Gamma_{i,l} (k_x,k_y)\, \tilde{M}_z^{(l)}(k_x,k_y)\\
    \tilde{M}_z^{(l)}(k_x,k_y) &= \sum_{i=1}^{N_l} [\Gamma^{-1}(k_x,k_y)]_{l,i} \, \tilde{B}_z(k_x,k_y,h_i)
\end{align}
with $\Gamma_{i,l} (k_x,k_y) = \tilde{K}_l (k_x,k_y,h_i)$. Finally, using reverse propagation method, we can record the change in layer-resolved magnetization in 3D Chern insulator slabs after applying a non-disruptive out-of-plane electric field via gating. The quantization rule Eq.~\eqref{alpha_zz_decoupled} of OME coupling can be then measured by finite difference.

\section{Discussions}

In real materials, each Chern layer has a finite thickness, so that we need to replace the interlayer distance by the lattice constant in the $z$ direction in all the results above. In particular, our proof by the Streda formula of the quantization rule Eq.~\eqref{alpha_zz_decoupled} is still valid since it relies only on the translation symmetry and the quantization of layer Chern number. We also note that Seleznev et al. \cite{seleznez-surfaceM-prb-2023} have recently proposed a new definition for the local marker of orbital magnetization, leading to another definition of layer-resolved orbital magnetization that differs from the one adopted in this work. Nevertheless, we have checked that our proposed quantization rule remains valid under this alternative definition of layer-projected orbital magnetization (see Sec.~\uppercase\expandafter{\romannumeral 9} in Supplemental Materials).

In fact, by adopting the ``gauge-discontinuity" formalism of Chern-Simons OME coupling \cite{liu-gaugeCS-prb-2015}, the quantization rule introduced in this work can be derived in a mathematically rigorous way even in the presence of considerable interlayer coupling. It can be interpreted as a new type of ``anomalous" Chern-Simons OME coupling unique to 3D Chern insulators. We refer the readers to Ref.~\onlinecite{xue-arxiv25} for the full mathematical derivations.

Furthermore, in genuine 3D Chern insulator, it worth noting that $\alpha_{zz}^{(l)}$ is only defined modulo $C e^2/h$, which is rooted in the indeterminacy of electric polarization that the choice of unit cell is arbitrary. 
Nevertheless, the difference between $\alpha_{zz}^{(l)}$ of two neighboring layers is physically measurable and is a topological quantity. The ambiguity in the branch choice of $\alpha_{zz}^{(l)}$ is either fixed by the symmetry constraint on the total bulk response $\sum_l \alpha^{(l)}$ or by boundary condition at the surface. We also note that when the total number of layers increases to a certain extent, even a very small electric field applied in the $z$ direction will cause Zener tunnelling and the system then is no longer an insulator. This imposes a truncation to the otherwise diverging layer-resolved OME response.

In summary, we propose a never-mentioned, quantized layer-resolved OME response in generic 3D Chern insulators. This OME response is quantized or half-quantized to $e^2/h$ in the presence of inversion or mirror symmetry. Furthermore, even lack of symmetries, the gradient of this OME coupling is still exactly quantized in unit of $e^2/h$ given by Eq.~\eqref{eq:quantization_rule_partial}. We further prove the quantization rule using the Streda formula, combined with numerical evidence that shows the robustness of the quantization rule against substrate effects and disorder. Through extensive first principles calculations, we propose that such considerable and quantized difference between layer-resolved OME is in principle measurable in Cr-doped $\rm{(Bi/Sb)}_2 \rm{Te}_3$ \cite{zhao-highCQAHE-nature-2020} and $\rm{MnBi}_2 \rm{Te}_4$ slabs. We also propose an detailed experimental workflow of measuring such quantization rule using state-of-the-art magnetometry by reverse propagation.

%\section{Data availability}
%The data that support the findings of this study are available from the corresponding author upon reasonable request.

%\section{Acknowledgement}
\acknowledgments
This work is supported by the National Natural Science Foundation of China (grant No. 12174257), the National Key Research and Development Program of China (grant No. 2024YFA1410400 and 2020YFA0309601). We also thank Haibiao Zhou and Weifeng Zhi for helpful discussions.

%\section{Author Contributions}
%J. L. conceived the idea and supervised the project. X. L., R. J., and Z. G. performed calculations. X. L. and J. L. wrote the manuscript with inputs from all authors. 

%\section{Competing Interests}
%The Authors declare no competing interests.

%\newpage
\bibliography{reference}

%%%%%%%%%%%%%%%%%%%%%%%%%%%%%%
%           SI               %
%%%%%%%%%%%%%%%%%%%%%%%%%%%%%%

\clearpage
\begin{widetext}

\begin{center}
\textbf{\large Supplemental Materials of ``Orbital magnetoelectric coupling of three dimensional Chern Insulator''}
\end{center}

\maketitle

%%%%%%%%%% Prefix a "S" to all equations, figures, tables and reset the counter %%%%%%%%%%
\setcounter{equation}{0}
\setcounter{figure}{0}
\setcounter{table}{0}
\setcounter{section}{0}
\setcounter{page}{1}
\makeatletter
\renewcommand{\theequation}{S\arabic{equation}}
\renewcommand{\thefigure}{S\arabic{figure}}
\renewcommand{\thetable}{S\arabic{table}}
% \renewcommand{\bibnumfmt}[1]{[S#1]}
% \newcommand{\setlabel}[1]{\edef\@currentlabel{#1}\label}
% \renewcommand{\citenumfont}[1]{S#1}
%%%%%%%%% Prefix a "S" to all equations, figures, tables and reset the counter %%%%%%%%%%

\section{Expressions of layer-resolved orbital magnetization and layer Chern number}
Our definition of layer-resolved orbital magnetization is inspired from the existing definition of local orbital magnetization given by \cite{resta-localM-prl13, resta-orbitalM-prb16}. In this section, we will derive several expressions for layer-resolved orbital magnetization in a slab. We also define layer Chern number using local Chern marker at the end of this section.

\subsection{Naive definition of layer-resolved orbital magnetization: first approach}
From classical electrodynamics, the orbital magnetization of a macroscopically homogeneous system is given by
\begin{equation}
    \label{eq:M_def_classical}
    \M = \frac{\m}{V}
    = \frac{1}{2V}\int  {\rm d}\vr \, \vr\times\textbf{j}(\vr),
\end{equation}
where $V$ is the volume of the system, $\vr$ is the position and $\textbf{j}$ is the current density. Note that the magnetic moment $\m$ is extensive while the magnetization $\M$ is intensive. In the thermodynamic limit with periodic boundary condition (PBC), a quantum expression for orbital $\M$ in 2D case, namely $M_z$, can be written as
\begin{equation}
\label{eqs:M_def_brute}
\begin{aligned}
    M_z &= \frac{-e}{2A}\sum_{E_n<\mu} \bra{\phi_n}\vr\times\textbf{v}\ket{\phi_n}\bigg|_z = -\frac{ie}{2\hbar A}\sum_{E_n<\mu}\bra{\phi_n}\vr\times[H,\vr]\ket{\phi_n}\bigg|_z \\
    &= -\frac{i e}{2\hbar A}\sum_{E_n<\mu}(\bra{\phi_n}xHy\ket{\phi_n}-\bra{\phi_n}yHx\ket{\phi_n})\\
    &= -\frac{ie}{2\hbar A}\sum_{E_n<\mu}2i\,{\rm Im}\bra{\phi_n}xHy\ket{\phi_n} = \frac{e}{\hbar A}{\rm Im}\sum_{E_n<\mu}\bra{\phi_n}xHy\ket{\phi_n}\\
    &\equiv \frac{e}{\hbar A}{\rm Im\,Tr}\{PxHyP\},
\end{aligned}
\end{equation}
where $H$ is the system's Hamiltonian, $A$ is the total area, $e>0$ is the elementary charge, and $\ket{\phi_n}$s are the eigenstates. Here, only spinless case is discussed. The first equality comes from $\textbf{v}=\frac{i}{\hbar}[H,\vr]$. The last equality use the occupied-state projector $P$. In the following, we also need the empty-state projector $Q$, the complement of $P$. So, we give their definition together 
\begin{equation}
    \label{eq:projector}
    P = \sum_{E_n<\mu}\ket{\phi_n}\bra{\phi_n},\quad Q = 1 - P.  
\end{equation}
The trace operation ${\rm Tr}$ is to sum over all the eigenstates of the Hilbert space.

Note that this expression only applies to systems with open boundary condition (OBC). This is because position operators are ill-defined in the thermodynamic limit and terms like $\bra{\phi_n}\vr\ket{\phi_n}$ diverge. It contains both the contribution of circulation of the current density from the bulk and the contribution from the surface. Then, one can define layer-resolved orbital magnetization as 
\begin{equation}
    \label{eqs:Mzl_def_brute}
    \begin{aligned}
        M_z^{(l)} = \frac{e}{\hbar A} {\rm Im} \sum_{\RR,s} \bra{s l,\RR} PxHyP \ket{s l,\RR},
    \end{aligned}
\end{equation}
where the atomic lattice basis $\ket{sl,\RR}$ represents the site at layer $l$, sublattice $s$ of unit-cell $\RR$. This formula is particularly useful when we introduce disorder in the system in OBC.

\subsection{Definition of layer-resolved orbital magnetization in the thermodynamic limit: second approach}
In the thermodynamic limit namely with PBC in the $x,y$-plane, we can derive another expression for $M_z^{(l)}$ \cite{resta-localM-prl13} starting from Eq.~\eqref{eqs:M_def_brute}. Since
\begin{equation}
    H = (P+Q)H(P+Q) = PHP + QHQ,
\end{equation}
then
\begin{equation}
\begin{aligned}
    M_z &= \frac{e}{\hbar A}{\rm Im\,Tr}\{Px(PHP+QHQ)yP\} = \frac{e}{\hbar A}{\rm Im\,Tr}\{PxPHPyP+PxQHQyP\}\\
    &=\frac{e}{\hbar A}{\rm Im\,Tr}\{PxQHQyP+(1-Q)xPHPy(1-Q)\},
\end{aligned}
\end{equation}
where
\begin{equation}
\begin{aligned}
    {\rm Im\,Tr}\{(1-Q)xPHPy(1-Q)\} &= {\rm Im\,Tr}\{xPHPy-QxPHPy-xPHPyQ+QxPHPyQ\}\\
    &={\rm Im\,Tr}\{-2QxPHPy(P+Q)+QxPHPyQ\} \\
    &= {\rm Im\,Tr}\{-2QxPHPyP-QxPHPyQ\}\\
    &={\rm Im\,Tr}\{-2PQxPHPy-QxPHPyQ\}\\
    &= {\rm Im\,Tr}\{-QxPHPyQ\},
\end{aligned}
\end{equation}
Here we use the fact that ${\rm Im\,Tr}\{xPHPy\}=0$, since 
\begin{equation}
    ({\rm Tr}\{xPHPy\})^* = {\rm Tr}\{yPHPx\} = {\rm Tr}\{xyPHP\} = {\rm Tr}\{yxPHP\} = {\rm Tr}\{xPHPy\}\,. 
\end{equation}
Finally, we get
\begin{equation}
\label{eq:Mz_Resta}
    M_z = \frac{e}{\hbar A}{\rm Im\,Tr}\{PxQHQyP-QxPHPyQ\}.
\end{equation}
or more generally,
\begin{equation}
    \label{eq:M_eps_Resta}
        M_\gamma = \frac{e}{2\hbar A}{\rm Im} [ \epsilon_{\gamma \alpha \beta}\,{\rm Tr} \{P r_\alpha QHQ r_\beta P - Qr_\alpha PHP r_\beta Q\} ]
\end{equation}
where the summation over repeated indices is implied and $\epsilon_{\gamma \alpha \beta}$ is the anti-symmetric tensor with $\alpha, \beta, \gamma = x,y,z$.

The key observation is that $P\vr Q$ and $Q\vr P$ are well-defined and regular even in an unbounded system within PBC. In principle, Eq.~\eqref{eq:Mz_Resta} only applies to a system which remains gapped both in bulk and at edges (since we assume $PHQ = QHP = 0$), and therefore does not apply, as such, to Chern insulators. However, we can make the substitution of $H\rightarrow H-\mu$ for Chern insulators or metals \cite{resta-localM-prl13}. Under this substitution, Eq.~\eqref{eq:Mz_Resta} is suitable to both systems with PBC and those with OBC.

% and the quantum expression of orbital magnetization \cite{resta-orbitalM-prl05, resta-orbitalM-prb06}
% \begin{equation}
% \label{M_kspace}
%     M_\gamma = -\frac{ie}{2\hbar}\epsilon_{\gamma\alpha\beta}\sum\limits_{E_{n\k }<\mu}\int_{BZ}(d\k )\bra{\partial_\alpha u_{n\k }}H_{\k }+E_{n\k }-2\mu\ket{\partial_\beta u_{n\k }}.
% \end{equation}
% where $(d\k ) = d^dk/(2\pi)^d$, $H_{\k } \ket{u_{n\k }} = E_{n\k }\ket{u_{n\k }}$, and $\mu$ is the chemical potential.

We derive the k-space expression \cite{resta-orbitalM-prb16} of Eq.~\eqref{eq:Mz_Resta} in a slab with PBC in the $x,y$ plane but finite in the $z$ direction. Suppose the eigenstates of the slab are $\ket{\psi_{n \k}}$. First, we need to evaluate
\begin{equation}
\begin{aligned}
    \bra{\psi_{n' \k '}} \vr \ket{\psi_{n\k }} &=\int {\rm d}\vr\,e^{-i\k '\cdot \vr } u_{n'\k '}^*(\vr)\, \vr\, e^{i\k \cdot\vr}u_{n\k }(\vr) \\
    &=\sum\limits_\RR \big(\int_{\Omega_0} d\vr\,e^{-i\k '\cdot\vr}u_{n'\k '}^*(\vr)(-i\partial_\k e^{i\k \cdot\vr})u_{n\k }(\vr) \big)e^{i(\k -\k ')\cdot\RR }\\
    &=N_{{\rm cell}}\,\delta_{\k ,\k '}\int_{\Omega_0}d\vr\,u_{n'\k }^*(\vr)i\,\partial_\k u_{n\k }\\
    &=N_{{\rm cell}}\,\delta_{\k ,\k '}\, i\braket{u_{n'\k }}{\partial_\k u_{n\k }}_{\Omega_0},
\end{aligned}
\end{equation}
where the third equality comes from $\partial_\k \braket{\psi_{n'\k }}{\psi_{n\k }} = 0$ and $N_{{\rm cell}}$ is the total number of unit cells. 
Here we choose to normalize the eigenstates over the unit cell's area $\Omega_0$, so $\braket{\psi_{n\k }}{\psi_{n\k }} = N_{{\rm cell}}$, while the definition of projectors is normalized to unity ($P^2 = P$). Therefore, we can express the projectors as
\begin{equation}
\begin{aligned}
    Q\vr P &= \frac{1}{N_{\rm cell}}\sum_{\k ,\,n\in{\rm occ.}}\frac{1}{N_{\rm cell}}\sum_{\k ',\,n'\in{\rm emp.}}\ket{\psi_{n'\k '}}\bra{\psi_{n'\k '}}\vr\ket{\psi_{n\k }}\bra{\psi_{n\k }}\\
    &= \frac{1}{N_{\rm cell}}\sum_{n\in{\rm occ}}\sum_{n'\in{\rm emp.}}\sum_{\k }\ket{\psi_{n'\k }}\,i\braket{u_{n'\k }}{\partial_\k u_{n\k }}_{\Omega_0}\bra{\psi_{n\k }} \,.
\end{aligned}
\end{equation}
where ``occ.'' and ``emp.'' represent indices of occupied and empty bands, respectively. Similarly,
\begin{equation}
    P\vr Q = \frac{1}{N_{\rm cell}}\sum_{n\in{\rm occ.}}\sum_{n'\in{\rm emp.}}\sum_{\k }\ket{\psi_{n\k }}\,i\braket{u_{n\k }}{\partial_\k u_{n'\k }}_{\Omega_0}\bra{\psi_{n'\k }}\,.
\end{equation}
Then,
\begin{equation}
\label{eq:PrQHQrP}
\begin{aligned}
    Pr_\alpha QHQr_\beta P&= \frac{1}{N_{\rm cell}^2}\sum_{\k ,\k '}\sum_{\substack{n\in{\rm occ.},\\ m\in{\rm emp.}}} \sum_{\substack{n'\in{\rm occ.},\\m'\in{\rm emp.}}} \ket{\psi_{n\k }}\,i\braket{u_{n\k }}{\partial_{k_\alpha}u_{m\k }}_{\Omega_0}\bra{\psi_{m\k }}H\ket{\psi_{m'\k '}}\,i\braket{u_{m'\k '}}{\partial_{k'_\beta}u_{n'\k '}}_{\Omega_0}\bra{\psi_{n'\k '}}\\
    &=\frac{1}{N_{\rm cell}}\sum_\k \sum_{\substack{n,n'\\ \in{\rm occ.}}}\sum_{\substack{m,m'\\ \in{\rm emp}}} \ket{\psi_{n\k }}\bra{\psi_{n'\k }}\braket{\partial_{k_\alpha}u_{n\k }}{u_{m\k }}_{\Omega_0} H_{mm'}(\k) \braket{u_{m'\k }}{\partial_{k_\beta}u_{n'\k }}_{\Omega_0},
\end{aligned}
\end{equation}
where $H_{mm'}(\k ) = \bra{u_{m\k }}H_\k \ket{u_{m'\k }}_{\Omega_0}$ and $H_\k  = e^{-i\k \cdot\vr}He^{i\k \vr}$. The second equality comes from $\bra{\psi_{m\k }}H\ket{\psi_{m'\k '}} = N_{\rm cell}\,\delta_{\k ,\k '}\bra{u_{m\k }}H_\k \ket{u_{m'\k '}}_{\Omega_0}$ and $\bra{u_{n\k }}\partial_{k_\alpha}u_{m\k }\rangle_{\Omega_0} = -\bra{\partial_{k_\alpha}u_{n\k }}u_{m\k }\rangle_{\Omega_0}$ since $\partial_{k_\alpha}\bra{u_{n\k }}u_{m\k }\rangle_{\Omega_0} = 0$. 

With anti-symmetric tensor, it is convenient to write
\begin{equation}
\begin{aligned}
    \epsilon_{\gamma\alpha\beta}Pr_\alpha QHQr_\beta P &= \epsilon_{\gamma\alpha\beta}\frac{1}{N_{\rm cell}}\sum_{\k }\sum_{n,n'\in{\rm occ.}}\big(\sum_{m,m'\in{\rm all}}-\sum_{m,m'\in{\rm occ.}}\big)\\
    &\ket{\psi_{n\k }}\bra{\psi_{n'\k }}\bra{\partial_{k_\alpha}u_{n\k }}u_{m\k }\rangle_{\Omega_0}H_{mm'}(\k)\bra{u_{m'\k }}\partial_{k_\beta}u_{n'\k }\rangle_{\Omega_0},
\end{aligned}
\end{equation}
where the second term in the parenthesis is vanishing by $\epsilon_{\gamma\alpha\beta}$ because of interchangeable indices. So,
\begin{equation}
\label{eq:eps_Tr_PrQHQrP}
\begin{aligned}
    \epsilon_{\gamma\alpha\beta}{\rm Tr}\{Pr_\alpha QHQr_\beta P\}&=\epsilon_{\gamma\alpha\beta}\sum_{\k }\sum_{n\in {\rm occ.}}\sum_{m\in{\rm all}}\bra{\partial_{k_\alpha}u_{n\k }}u_{m\k }\rangle_{\Omega_0}E_m\bra{u_{m\k }}\partial_{k_\beta}u_{n\k }\rangle_{\Omega_0}\\&=\epsilon_{\gamma\alpha\beta}\sum_{\k }\sum_{n\in{\rm occ.}}\bra{\partial_{k_\alpha}u_{n\k }}H_\k \ket{\partial_{k_\beta}u_{n\k }}_{\Omega_0}.
\end{aligned}
\end{equation}
Similarly,
\begin{equation}
\label{eq:eps_Tr_QrPHPrQ}
\begin{aligned}
    &\epsilon_{\gamma\alpha\beta}{\rm Tr}\{Qr_\alpha PHPr_\beta Q\}\\
    &=\epsilon_{\gamma\alpha\beta}\frac{1}{N_{\rm cell}^2}{\rm Tr}\big \{\sum_{\k ,\k '}\sum_{n,n'\in{\rm occ.}}\sum_{m,m'\in{\rm emp.}}\ket{\psi_{m\k }}i\bra{u_{m\k }}\partial_{k_\alpha}u_{n\k }\rangle_{\Omega_0}\bra{\psi_{n\k }}H\ket{\psi_{n'\k '}}i\bra{u_{n'\k '}}\partial_{k_\beta}u_{m'\k '}\rangle_{\Omega_0}\bra{\psi_{m'\k }}\big \}\\
    &= \epsilon_{\gamma\alpha\beta}\sum_\k \sum_{n\in{\rm occ.}}\sum_{m\in {\rm emp.}}\bra{\partial_{k_\alpha}u_{m\k }}u_{n\k }\rangle_{\Omega_0}\bra{u_{n\k }}H_\k \ket{u_{n\k }}_{\Omega_0}\bra{u_{n\k }}\partial_{k_\beta}u_{m\k }\rangle_{\Omega_0}\\
    &=\epsilon_{\gamma\alpha\beta}\sum_\k \sum_{n\in{\rm occ.}}\sum_{m\in {\rm all}} E_{n\k }\bra{\partial_{k_\beta}u_{n\k }}u_{m\k }\rangle_{\Omega_0}\bra{u_{m\k }}\partial_{k_\alpha}u_{n\k }\rangle_{\Omega_0}\\
    &=\epsilon_{\gamma\alpha\beta}\sum_\k \sum_{n\in{\rm occ.}}\bra{\partial_{k_\beta}u_{n\k }}\partial_{k_\alpha}u_{n\k }\rangle E_{n\k }
    = -\epsilon_{\gamma\alpha\beta}\sum_\k \sum_{n\in{\rm occ.}}\bra{\partial_{k_\alpha}u_{n\k }}\partial_{k_\beta}u_{n\k }\rangle E_{n\k }.
\end{aligned}
\end{equation}
% So, Eq.(\ref{m_OBC_Resta}) becomes
% \begin{equation}
% \label{m_bulk}
% \begin{aligned}
%     m_\gamma &= \frac{e}{2\hbar}{\rm Im}\{\epsilon_{\gamma\alpha\beta}\sum_{\k \in{\rm BZ}}\sum_{n\in{\rm occ}}\bra{\partial_{k_\alpha}u_{n\k }}H_\k +E_{n\k }\ket{\partial_{k_\beta}u_{n\k }}\}\\
%     &=\frac{e}{2\hbar}{\rm Im}\{\epsilon_{\gamma\alpha\beta}\int\frac{d^d\k }{(2\pi)^d}\,\Omega_0 N_{\rm cell} \sum_{n\in{\rm occ}}\bra{\partial_{k_\alpha}u_{n\k }}H_\k +E_{n\k }\ket{\partial_{k_\beta}u_{n\k }}\\
%     &= M^{(bulk)}\, \Omega_0 N_{\rm cell},
% \end{aligned}
% \end{equation}
% $M^{(bulk)}$ is the bulk contribution after comparing with Eq.(\ref{M_kspace}). The rest of it is then the surface contribution, which is actually the total Chern number of the filled bands. 

Suppose the slab consists of $N_{\L}$ layers in the $z$ direction. The Bloch function can be written as
\begin{equation}
\label{eq:psi_layer_resolved}
    \ket{\psi_{n\k }} = \sum_s\sum_{l=1}^{N_{\rm L}}C_{s l,n}(\k )\ket{s l,\k } \quad {\rm with} \quad \k  = (k_x, k_y),
\end{equation}
where the Fourier basis $\ket{s l,\k }$ is defined as 
\begin{equation}
\left\{
\begin{aligned}
    \ket{s l,\k } &= \sum_\RR e^{i\k \cdot\RR } \ket{s l,\RR },\\
    \ket{s l,\RR } &=\frac{1}{N_{\rm cell}}\sum_\k e^{-i\k \cdot\RR }\ket{s l,\k },
\end{aligned}  
\right.
\end{equation}
where $s$ means sublattice and $l$ is layer index. The corresponding normalization relation becomes
\begin{equation}
    \left\{
\begin{aligned}
    \braket{s l,\RR }{s'l',\RR '} &= \delta_{ss'}\delta_{ll'}\delta_{\RR \RR '},\\
    \braket{s l,\k }{s'l',\k '} &= N_{\rm cell}\delta_{ss'}\delta_{ll'}\delta_{\k \k '},\\
    \braket{\psi_{n\k }}{\psi_{n'\k '}}&= N_{\rm cell}\delta_{nn'}\delta_{\k \k '}\,.
\end{aligned}  
\right.
\end{equation}
% From Eq.(\ref{PrQHQrP}), we know that
% \begin{equation}
% \begin{aligned}
%     Pr_\alpha QHQr_\beta P =& \frac{1}{N_{\rm cell}}\sum_{\k }\sum_{n,n'\in{\rm occ}}\sum_{m\in{\rm emp}}E_{m\k }\bra{\partial_{k_\alpha}u_{n\k }}u_{m\k }\rangle_{\Omega_0}\bra{u_{m\k }}\partial_{k_\beta}u_{n'\k }\rangle_{\Omega_0}\\&\sum_{s l,s' l'}C_{s l,n}(\k )C_{s'l',n'}^*(\k )\ket{s l,\k }\bra{s'l',\k }.
% \end{aligned}
% \end{equation}
So, from Eq.~\eqref{eq:eps_Tr_PrQHQrP}, we get
\begin{equation}
\begin{aligned}
    \epsilon_{\gamma\alpha\beta}&{\rm Tr}\{Pr_\alpha QHQr_\beta P\} = \epsilon_{\gamma\alpha\beta}\sum_\k \sum_{n\in{\rm occ.}}\bra{\partial_{k_x}u_{n\k }}H_\k \ket{\partial_{k_\beta}u_{n\k }}_{\Omega_0}\\&=\epsilon_{\gamma\alpha\beta}\sum_\k \sum_{n,n'\in{\rm occ.}}\sum_{m\in{\rm emp.}}\sum_{s\,l}C_{s l,n}(\k )C_{s l,n'}^*(\k )E_{m\k }\bra{\partial_{k_\alpha}u_{n\k }}u_{m\k }\rangle_{\Omega_0}\bra{u_{m\k }}\partial_{k_\beta}u_{n'\k }\rangle_{\Omega_0}.
\end{aligned}
\end{equation}
Similarly, Eq.~\eqref{eq:eps_Tr_QrPHPrQ} becomes
\begin{equation}
\begin{aligned}
    \epsilon_{\gamma\alpha\beta}&{\rm Tr}\{Qr_\alpha PHPr_\beta Q\} = -\epsilon_{\gamma\alpha\beta}\sum_\k \sum_{n\in{\rm occ.}}\bra{\partial_{k_\alpha}u_{n\k }}\partial_{k_\beta}u_{n\k }\rangle E_{n\k }\\&=\epsilon_{\gamma\alpha\beta}\sum_\k \sum_{n\in{\rm occ.}}\sum_{m,m'\in{\rm emp.}}\sum_{s\,l}C_{s l,m}(\k )C_{s l,m'}^*(\k )E_{n\k }\bra{u_{m\k }}\partial_{k_\alpha}u_{n\k }\rangle_{\Omega_0}\bra{\partial_{k_\beta}u_{n\k }}u_{m'\k }\rangle_{\Omega_0}.
\end{aligned}
\end{equation}
So, $M_z$ for the slab reads:
\begin{equation}
\label{eq:Mzl_def_k}
\begin{aligned}
    M_z &= \frac{e}{2\hbar A}{\rm Im}\big \{\epsilon_{z\alpha\beta}\sum_{s\,l}\sum_\k \sum_{n,n'\in{\rm occ.}}\sum_{m\in{\rm emp.}}C_{s l,n}(\k )C_{s l,n'}^*(\k )E_{m\k }\bra{u_{n\k }}\partial_{k_\alpha}u_{m\k }\rangle_{\Omega_0}\bra{\partial_{k_\beta}u_{m\k }}u_{n'\k }\rangle_{\Omega_0}\big \}\\
    &-\frac{e}{2\hbar A}{\rm Im}\big \{\epsilon_{z\alpha\beta}\sum_{s\,l}\sum_\k \sum_{n\in{\rm occ.}}\sum_{m,m'\in{\rm emp.}}C_{s l,m}(\k )C_{s l,m'}^*(\k )E_{n\k }\bra{u_{m\k }}\partial_{k_\alpha}u_{n\k }\rangle_{\Omega_0}\bra{\partial_{k_\beta}u_{n\k }}u_{m'\k }\rangle_{\Omega_0}\big \}\\
    &\equiv\sum_{l}M_z^{(l)},
\end{aligned}
\end{equation}
where the layer-resolved orbital magnetization of the $l$-th layer $M_z^{(l)}$ is defined by summing all the degrees of freedom except layer $l$. As discussed before, when considering Chern insulators having surface contribution, one needs to replace $H\rightarrow H-\mu$ and $E_{n\k }\rightarrow E_{n\k }-\mu$ in the previous formula.

\subsection{Layer Chern number}
In the same spirit, we can also define layer Chern number for a slab using local Chern marker \cite{resta-orbitalM-prb16,resta-localchern-prb11}
\begin{equation}
    C_\gamma(\vr) = \frac{i}{4\pi}\epsilon_{\gamma\alpha\beta}\bra{\vr}Pr_\alpha Qr_\beta P-Qr_\alpha Pr_\beta Q\ket{\vr}.
\end{equation}
In the Fourier basis, we can write down the k-space expression for layer Chern number
\begin{equation}
    \label{eqs:layer_chern}
\begin{aligned}
    C_z &= -\frac{1}{4 \pi}\epsilon_{z\alpha\beta}\,{\rm Im \, Tr}\{Pr_\alpha Qr_\beta P-Qr_\alpha Pr_\beta Q \}\\
    &=-\frac{\pi}{A}\epsilon_{z\alpha\beta}\,{\rm Im} \left\{\sum_{s\,l}\sum_{n,n'\in{\rm occ}}\sum_{m\in{\rm emp}}\sum_{\k }C_{sl,n}(\k )C_{sl,n'}^*(\k )\bra{u_{n\k }}\partial_{k_\alpha}u_{m\k }\rangle_{\Omega_0}\bra{\partial_{k_\beta}u_{m\k }}u_{n'\k }\rangle_{\Omega_0} \right.\\
    & \left. \quad \quad -\sum_{s\,l}\sum_{n\in{\rm occ}}\sum_{m,m'\in{\rm emp}}\sum_{\k }C_{sl,m}(\k )C_{sl,m'}^*(\k )\bra{u_{m\k }}\partial_{k_\alpha}u_{n\k }\rangle_{\Omega_0}\bra{\partial_{k_\beta}u_{n\k }}u_{m'\k }\rangle_{\Omega_0} \right\}\\
    &\equiv\sum_{l}C_z(l).
\end{aligned}
\end{equation}

\section{Layer-resolved orbital magnetization under finite vertical electric field}

We show in Fig.~\ref{fig:alpha_PBC} that $M_z^{(l)}$ varies linearly with weak electric field for both types of stacking, independent of number of layers. Here, we calculate $M_z^{(l)}$ using the definition Eq.~\eqref{eq:Mzl_def_k}.
\begin{figure}[h]
   \includegraphics[width=0.5\textwidth]{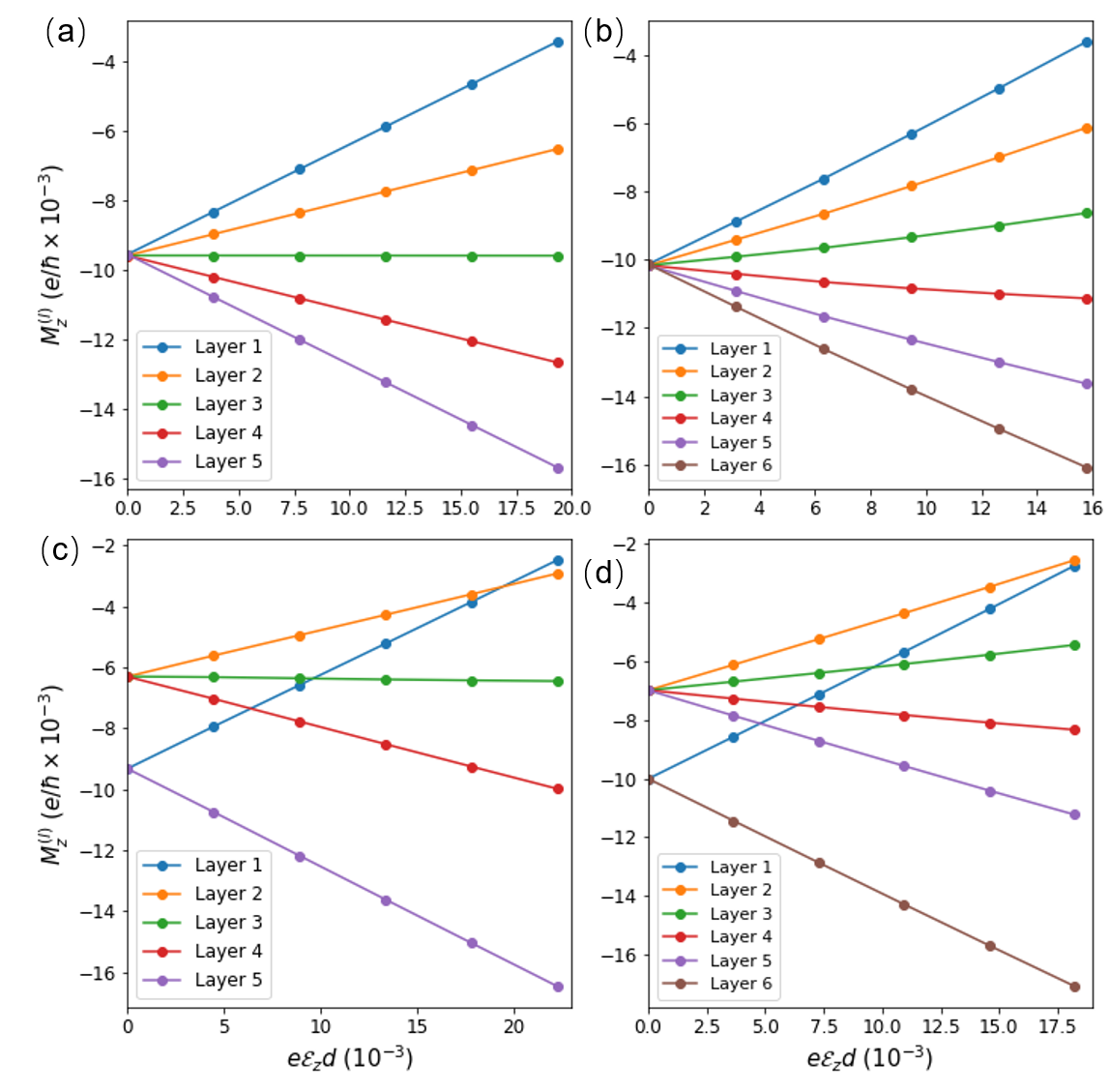}
   \caption{Layer-resolved $M_z^{(l)}$ of the slab under several $\mathcal{E}_z$ for different stacking ways and total layer numbers: (a) AA stacking, 5 layers; (b) AA stacking, 6 layers; (c) AB stacking, 5 layers; (d) AB stacking, 6 layers.}
   \label{fig:alpha_PBC}
\end{figure}

\section{Perturbation theory for layer-resolved orbital magnetoelectric response}
When a vertical electric field is applied to a slab, our perturbed Hamiltonian reads: $H = H_0 +e\mathcal{E} z$. Suppose the electric field is so weak such that bulk gap is not closed and the potential drop between the top and bottom layer is the smallest energy scale in the system. The idea is to express Eq.~\eqref{eq:Mzl_def_k} of perturbed Hamiltonian $H$ using the information of $H_0$. By singling out the terms proportional to $\mathcal{E}$, the layer-resolved orbital magnetoelectric response $\alpha_{zz}^{(l)}$ is then derived by $\partial M_z^{(l)} / \partial \mathcal{E}$. After gathering all the first-order corrections (including energy, wavefunction and chemical potential with particle number fixed), a semi-analytic formula for $\alpha_{zz}^{(l)}$ can be derived. We say ``semi-analytic'' since the determination of the chemical potential still requires solving numerically the system with open boundary condition, which is unavoidable due to the presence of gapless surface states in 3D Chern insulators. The detailed derivation below can be skipped if the readers are not interested in long but direct algebra. The derived formula is benchmarked to be correct as evidenced by finite-difference method as shown in Table~\I\  in the main text and also Table~\ref{tab:layer_resolved_PBC}.

Let us first focus on the first-order correction in $\mathcal{E}$ to all terms in Eq.~\eqref{eq:Mzl_def_k}. First, the correction in wavefunction reads
\begin{equation}
\label{eq:pert_u_nk}
    \begin{aligned}
        \ket{u_{n\k }} &= \ket{u_{n\k }^{(0)}}+\sum_{n'\neq n}\frac{\frac{1}{N_{\rm cell}}\bra{u_{n'\k }^{(0)}}e \mathcal{E} z\ket{u_{n\k }^{(0)}}}{E_{n\k }^{(0)}-E_{n'\k }^{(0)}}\ket{u_{n'\k}^{(0)}}\\
        &=\ket{u_{n\k }^{(0)}}+\sum_{n\neq n'}\frac{\frac{1}{N_{\rm cell}}\sum_{sl}\sum_{s'l'}C_{s'l',n'}^{(0)\,*}(\k )C_{sl,n}^{(0)}(\k )\bra{s'l',\k }e\mathcal{E} z\ket{sl,\k }}{E_{n\k }^{(0)}-E_{n'\k }^{(0)}}\ket{u_{n'\k }^{(0)}}\\
        &=\ket{u_{n\k }^{(0)}}+\sum_{n'\neq n}\sum_{sl}\frac{C_{sl,n'}^{(0)\,*}(\k )C_{sl,n}^{(0)}(\k )}{E_{n\k }^{(0)}-E_{n'\k }^{(0)}}\,e\mathcal{E} ld \ket{u_{n'\k }^{(0)}}\\
        &\equiv \ket{u_{n\k }^{(0)}}+\ket{u_{n\k }^{(1)}},
    \end{aligned}
\end{equation}
where the upper index ``$(0)$'' means the quantity defined for non-perturbed Hamiltonian $H_0$. The third line comes from $\bra{s'l',\k }e\mathcal{E} z\ket{sl,\k }=e\mathcal{E} ld\,\delta_{s'l',sl}N_{\rm cell}$, in which $l$ is the layer index and $d$ is the interlayer distance. 

The correction in the Fourier expansion reads
\begin{equation}
\label{eq:pert_C_sln}
    \begin{aligned}
        C_{sl,n}(\k )&\equiv\bra{sl,\k }u_{n\k }\rangle\\
        &= C_{sl,n}^{(0)}(\k )+\sum_{n'\neq n}\sum_{s'\,l'}\frac{C_{s'l',n'}^{(0)\,*}(\k )C_{s'l',n'}^{(0)}(\k )}{E_{n\k }^{(0)}-E_{n'\k }^{(0)}}\,e \mathcal{E} l'd\,C_{sl,n'}^{(0)}(\k )\\
        &\equiv C_{sl,n}^{(0)}(\k )+C_{sl,n}^{(1)}(\k ).
    \end{aligned}
\end{equation}

The derivatives of the wavefunction reads
\begin{equation}
\label{eq:pert_partial_k_u_nk}
    \ket{\partial_{k_\alpha}u_{n\k }} = -iA_{n,\alpha}(\k )\ket{u_{n\k }}+\sum_{m\neq n}\frac{\frac{1}{N_{\rm cell}}\ket{u_{m\k }}\bra{u_{m\k }}}{E_{n\k }-E_{m\k }}\,(\partial_{k_\alpha}H_\k )\ket{u_{n\k }},
\end{equation}
with Berry connection $A_{n,\alpha}(\k ) = i\bra{u_{n\k }}\partial_{k_\alpha}u_{n\k }\rangle$. Note from Eq.~\eqref{eq:Mzl_def_k} that we only need to calculate $\bra{u_{m\k }}\partial_{k_\alpha}u_{n\k }\rangle_{\Omega_0}$ with $m\neq n$ so we can omit the first term in Eq.~\eqref{eq:pert_partial_k_u_nk}. We also note that $\partial_{k_\alpha}H_\k  = \partial_{k_\alpha}H_\k ^{(0)}$ with $H_\k ^{(0)} = e^{-i\k \cdot\vr}H_0\,e^{i\k \cdot\vr}$. Therefore we need to single out the first-order correction to $\ket{\partial_{k_\alpha}u_{n\k }}$ by expanding all other terms. The simplest one would be $E_{n\k }$
\begin{equation}
    \begin{aligned}
        E_{n\k } &= E_{n\k }^{(0)}+\bra{u_{n\k }^{(0)}}e\mathcal{E} z\ket{u_{n\k }^{(0)}}\\
        &= E_{n\k }^{(0)} +\sum_{s\,l}C_{sl,n}^{(0) \, *}(\k )C_{sl,n}^{(0)}(\k )\,e\mathcal{E} ld\\
        &\equiv E_{n\k }^{(0)} + E_{n\k }^{(1)}.
    \end{aligned}
\end{equation}
Let's now expand the second term in Eq.~\eqref{eq:pert_partial_k_u_nk}. Since
\begin{equation*}
    E_{n\k }-E_{m\k } = E_{n\k }^{(0)}-E_{m\k }^{(0)}+E_{n\k }^{(1)}-E_{m\k }^{(1)},
\end{equation*}
\begin{equation*}
    \ket{u_{m\k }} = \ket{u_{m\k }^{(0)}}+\ket{u_{m\k }^{(1)}},
\end{equation*}
\begin{equation*}
    \bra{u_{m\k }}\partial_{k_\alpha}H_\k \ket{u_{n\k }} = \bra{u_{m\k }^{(0)}}\partial_{k_\alpha}H_\k ^{(0)}\ket{u_{n\k }^{(0)}} +\bra{u_{m\k }^{(1)}}\partial_{k_\alpha}H_\k ^{(0)}\ket{u_{n\k }^{(0)}}+\bra{u_{m\k }^{(0)}}\partial_{k_\alpha}H_\k ^{(0)}\ket{u_{n\k }^{(1)}},
\end{equation*}
so,
\begin{equation*}
\begin{aligned}
    \ket{u_{m\k }}\bra{u_{m\k }}\partial_{k_\alpha}H_\k \ket{u_{n\k }} =&\ket{u_{m\k }^{(0)}}\bra{u_{m\k }^{(0)}}\partial_{k_\alpha}H_\k ^{(0)}\ket{u_{n\k }^{(0)}}+\ket{u_{m\k }^{(1)}}\bra{u_{m\k }^{(0)}}\partial_{k_\alpha}H_\k ^{(0)}\ket{u_{n\k }^{(0)}}\\&+\ket{u_{m\k }^{(0)}}\Big(\bra{u_{m\k }^{(1)}}\partial_{k_\alpha}H_\k ^{(0)}\ket{u_{n\k }^{(0)}}+\bra{u_{m\k }^{(0)}}\partial_{k_\alpha}H_\k ^{(0)}\ket{u_{n\k }^{(1)}}\Big),
\end{aligned}
\end{equation*}
\begin{equation*}
    \frac{1}{E_{n\k }-E_{m\k }} = \frac{1}{E_{n\k }^{(0)}-E_{m\k }^{(0)}}-\frac{E_{n\k }^{(1)}-E_{m\k }^{(1)}}{\big(E_{n\k }^{(0)}-E_{m\k }^{(0)}\big)^2}.
\end{equation*}
Finally,
\begin{equation}
\begin{aligned}
    &\frac{1}{N_{\rm cell}}\sum_{m\neq n}\frac{\ket{u_{m\k }}\bra{u_{m\k }}\partial_{k_\alpha}H_\k \ket{u_{n\k }}}{E_{n\k }-E_{m\k }}\\
    &=\frac{1}{N_{\rm cell}}\sum_{m\neq n} \Big[\frac{\ket{u_{m\k }^{(0)}}\bra{u_{m\k }^{(0)}}v_\alpha\ket{u_{m\k }^{(0)}}}{E_{n\k }^{(0)}-E_{m\k }^{(0)}}-\frac{\ket{u_{m\k }^{(0)}}\bra{u_{m\k }^{(0)}}v_\alpha\ket{u_{m\k }^{(0)}}}{(E_{n\k }^{(0)}-E_{m\k }^{(0)})^2}(E_{n\k }^{(1)}-E_{m\k }^{(1)})\\&+\frac{\ket{u_{m\k }^{(1)}}\bra{u_{m\k }^{(0)}}v_\alpha\ket{u_{m\k }^{(0)}}}{E_{n\k }^{(0)}-E_{m\k }^{(0)}}+\frac{\ket{u_{m\k }^{(0)}}\big(\bra{u_{m\k }^{(1)}}v_\alpha\ket{u_{m\k }^{(0)}}+\bra{u_{m\k }^{(0)}}v_\alpha\ket{u_{m\k }^{(1)}}\big)}{E_{n\k }^{(0)}-E_{m\k }^{(0)}}
    \Big]\\&\equiv \ket{\partial_{k_\alpha}u_{n\k }^{(0)}}+\ket{\partial_{k_\alpha}u_{n\k }^{(1)}},
\end{aligned}
\end{equation}
where $\partial_{k_\alpha}H_\k  = \partial_{k_\alpha}H_\k ^{(0)}\equiv v_\alpha$. 

With all the ingredients derived above, we can expand Eq.~\eqref{eq:Mzl_def_k} to the first-order of $\mathcal{E}$ field. Typically, we need to treat terms like:
\begin{equation*}
    (*) = \underbrace{C_{sl,n}(\k )C_{sl,n'}^*(\k )}_{({\rm I})}E_{m\k }\underbrace{\bra{u_{n\k }}\partial_{k_\alpha}u_{m\k }\rangle_{\Omega_0}}_{({\rm II})}\bra{\partial_{k_\beta}u_{m\k }}u_{n'\k }\rangle_{\Omega_0},
\end{equation*}
\begin{equation*}
\begin{aligned}
    ({\rm I}) &= C_{sl,n}^{(0)}(\k )C_{sl,n'}^{(0)*}(\k )+C_{sl,n}^{(0)}(\k )C_{sl,n'}^{(1)*}(\k )+C_{sl,n}^{(1)}(\k )C_{sl,n'}^{(0)*}(\k ),\\
    ({\rm II}) &= (\bra{u_{n\k }^{(0)}}+\bra{u_{n\k }^{(1)}})(\ket{\partial_{k_\alpha}u_{m\k }^{(0)}}+\ket{\partial_{k_\alpha}u_{m\k }^{(1)}})\\
    &=\bra{u_{n\k }^{(0)}}\partial_{k_\alpha}u_{m\k }^{(0)}\rangle+\bra{u_{n\k }^{(0)}}\partial_{k_\alpha}u_{m\k }^{(1)}\rangle+\bra{u_{n\k }^{(1)}}\partial_{k_\alpha}u_{m\k }^{(0)}\rangle.
\end{aligned}
\end{equation*}
Then
\begin{equation}
    \label{eq:typical_term_star}
    \begin{aligned}
        (*) &= C_{sl,n}^{(0)}(\k )C_{sl,n'}^{(0)*}(\k )E_{m\k }^{(0)}\bra{u_{n\k }^{(0)}}\partial_{k_\alpha}u_{m\k }^{(0)}\rangle_{\Omega_0}\bra{\partial_{k_\beta}u_{m\k }^{(0)}}u_{n'\k }^{(0)}\rangle_{\Omega_0}\\
        &+\big(C_{sl,n}^{(1)}(\k )C_{sl,n'}^{(0)*}(\k )+C_{sl,n}^{(0)}(\k )C_{sl,n'}^{(1)*}(\k )\big)E_{m\k }^{(0)}\bra{u_{n\k }^{(0)}}\partial_{k_\alpha}u_{m\k }^{(0)}\rangle_{\Omega_0}\bra{\partial_{k_\beta}u_{m\k }^{(0)}}u_{n'\k }^{(0)}\rangle_{\Omega_0}\\
        &+C_{sl,n}^{(0)}(\k )C_{sl,n'}^{(0)*}(\k )E_{m\k }^{(1)}\bra{u_{n\k }^{(0)}}\partial_{k_\alpha}u_{m\k }^{(0)}\rangle_{\Omega_0}\bra{\partial_{k_\beta}u_{m\k }^{(0)}}u_{n'\k }^{(0)}\rangle_{\Omega_0}\\
        &+C_{sl,n}^{(0)}(\k )C_{sl,n'}^{(0)*}(\k )E_{m\k }^{(0)}\big(\bra{u_{n\k }^{(1)}}\partial_{k_\alpha}u_{m\k }^{(0)}\rangle_{\Omega_0}+\bra{u_{n\k }^{(0)}}\partial_{k_\alpha}u_{m\k }^{(1)}\rangle_{\Omega_0}\big)\bra{\partial_{k_\beta}u_{m\k }^{(0)}}u_{n'\k }^{(0)}\rangle_{\Omega_0}\\
        &+C_{sl,n}^{(0)}(\k )C_{sl,n'}^{(0)*}(\k )E_{m\k }^{(0)}\bra{u_{n\k }^{(0)}}\partial_{k_\alpha}u_{m\k }^{(0)}\rangle_{\Omega_0}\big(\bra{\partial_{k_\beta}u_{m\k }^{(1)}}u_{n'\k }^{(0)}\rangle_{\Omega_0}+\bra{\partial_{k_\beta}u_{m\k }^{(0)}}u_{n'\k }^{(1)}\rangle_{\Omega_0}\big).
    \end{aligned}
\end{equation}
Now we need to explicit $\bra{u_{n\k }^{(0)}}\partial_{k_\alpha}u_{m\k }^{(0)}\rangle$, $\bra{u_{n\k }^{(1)}}\partial_{k_\alpha}u_{m\k }^{(0)}\rangle$ and $\bra{u_{n\k }^{(0)}}\partial_{k_\alpha}u_{m\k }^{(1)}\rangle$. First,
\begin{equation}
\label{unk0_Pumk0}
    \bra{u_{n\k }^{(0)}}\partial_{k_\alpha}u_{m\k }^{(0)}\rangle_{\Omega_0}\xlongequal{m\neq n}\bra{u_{n\k }^{(0)}}\frac{1}{N_{\rm cell}}\sum_{m'\neq m}\frac{\ket{u_{m'\k }^{(0)}}\bra{u_{m'\k }^{(0)}}v_\alpha\ket{u_{m\k }^{(0)}}}{E_{m\k }^{(0)}-E_{m'\k }^{(0)}}=\frac{\bra{u_{n\k }^{(0)}}v_\alpha\ket{u_{m\k }^{(0)}}_{\Omega_0}}{E_{m\k }^{(0)}-E_{n\k }^{(0)}},
\end{equation}
where $\bra{u_{n\k }^{(0)}}u_{m'\k }^{(0)}\rangle = N_{\rm cell}\,\delta_{nm'}$.
\begin{equation}
\label{unk1_Pumk0}
\begin{aligned}
    \bra{u_{n\k }^{(1)}}\partial_{k_\alpha}u_{m\k }^{(0)}\rangle_{\Omega_0}&\xlongequal{m\neq n}\sum_{n'\neq n; s\,l}\frac{C_{sl,n}^{(0)*}(\k )C_{sl,n'}^{(0)}(\k )}{E_{n\k }^{(0)}-E_{n'\k }^{(0)}}\,e\mathcal{E} ld\bra{u_{n'\k }^{(0)}}\frac{1}{N_{\rm cell}}\sum_{m'\neq m}\frac{\ket{u_{m'\k }^{(0)}}\bra{u_{m'\k }^{(0)}}v_\alpha\ket{u_{m\k }^{(0)}}}{E_{m\k }^{(0)}-E_{m'\k }^{(0)}}\\
    &=\sum_{n'\neq n,m; s\,l}\frac{C_{sl,n}^{(0)*}(\k )C_{sl,n'}^{(0)}(\k )}{E_{n\k }^{(0)}-E_{n'\k }^{(0)}}\frac{\bra{u_{n'\k }^{(0)}}v_\alpha\ket{u_{m\k }^{(0)}}_{\Omega_0}}{E_{m\k }^{(0)}-E_{n'\k }^{(0)}}\,e\mathcal{E} ld.
\end{aligned}
\end{equation}
\begin{equation}
\label{unk0_Pumk1}
\begin{aligned}
    \bra{u_{n\k }^{(0)}}\partial_{k_\alpha}u_{m\k }^{(1)}\rangle_{\Omega_0}&\xlongequal{m\neq n}\bra{u_{n\k }^{(0)}}\frac{1}{N_{\rm cell}}\sum_{m'\neq m}\Big[-\frac{\ket{u_{m'\k }^{(0)}}\bra{u_{m'\k }^{(0)}}v_\alpha\ket{u_{m\k }^{(0)}}_{\Omega_0}}{(E_{m\k }^{(0)}-E_{m'\k }^{(0)})^2}(E_{m\k }^{(1)}-E_{m'\k }^{(1)})\\&+\frac{\ket{u_{m'\k }^{(1)}}\bra{u_{m'\k }^{(0)}}v_\alpha\ket{u_{m\k }^{(0)}}_{\Omega_0}}{E_{m\k }^{(0)}-E_{m'\k }^{(0)}}+\frac{\ket{u_{m'\k }^{(0)}}\big(\bra{u_{m'\k }^{(1)}}v_\alpha\ket{u_{m\k }^{(0)}}_{\Omega_0}+\bra{u_{m'\k }^{(0)}}v_\alpha\ket{u_{m\k }^{(1)}}_{\Omega_0}\big)}{E_{m\k }^{(0)}-E_{m'\k }^{(0)}}\Big]\\
    &=-\frac{\bra{u_{n\k }^{(0)}}v_\alpha\ket{u_{m\k }^{(0)}}_{\Omega_0}}{(E_{m\k }^{(0)}-E_{n\k }^{(0)})^2}(E_{m\k }^{(1)}-E_{n\k }^{(1)})+\sum_{m'\neq m}\frac{\bra{u_{n\k }^{(0)}}u_{m'\k }^{(1)}\rangle_{\Omega_0}\bra{u_{m'\k }^{(0)}}v_\alpha\ket{u_{m\k }^{(0)}}_{\Omega_0}}{E_{m\k }^{(0)}-E_{m'\k }^{(0)}}\\&+\frac{\bra{u_{n\k }^{(1)}}v_\alpha\ket{u_{m\k }^{(0)}}_{\Omega_0}+\bra{u_{n\k }^{(0)}}v_\alpha\ket{u_{m\k }^{(1)}}_{\Omega_0}}{E_{m\k }^{(0)}-E_{n\k }^{(0)}}.
\end{aligned}
\end{equation}
Now we further:
\begin{equation}
\begin{aligned}
    \bra{u_{n\k }^{(0)}}u_{m'\k }^{(1)}\rangle_{\Omega_0}&=\sum_{s\,l}\sum_{m''\neq m'}\bra{u_{n\k }^{(0)}}u_{m''\k }^{(0)}\rangle_{\Omega_0}\frac{C_{sl,m''}^{(0)*}(\k )C_{sl,m'}^{(0)}(\k )}{E_{m'\k }^{(0)}-E_{m''\k }^{(0)}}\,e\mathcal{E} ld\\
    &=\sum_{s\,l}\frac{C_{sl,n}^{(0)*}(\k )C_{sl,m'}^{(0)}(\k )}{E_{m'\k }^{(0)}-E_{n\k }^{(0)}}(1-\delta_{nm'})\,e\mathcal{E} ld,
\end{aligned}
\end{equation}
where $\bra{u_{n\k }^{(0)}}u_{m''\k }^{(0)}\rangle_{\Omega_0} = \delta_{nm''}$. 

For simplicity, we introduce the following shorthand notations:
\begin{equation}
\begin{aligned}
    v_{\alpha,nm}(\k )&\equiv\bra{u_{n\k }^{(0)}}v_\alpha\ket{u_{m\k }^{(0)}}_{\Omega_0},\\
    \Delta_{\k ,mn} &\equiv E_{m\k }^{(0)}-E_{n\k }^{(0)},\\
    \Tilde{\mathcal{E}}&\equiv e\mathcal{E} d,\\
    D_{sl,nm}(\k )&\equiv C_{sl,n}^{(0)*}(\k )C_{sl,m}^{(0)}(\k ),\\
    \Rightarrow D_{sl,nm}^*(\k )&=D_{sl,mn}(\k ),\quad v_{\alpha,nm}^*(\k ) = v_{\alpha,mn}(\k ).
\end{aligned}
\end{equation}
Then, for example, the expressions involving velocity operators can be rewritten as
\begin{equation}
\begin{aligned}
    \bra{u_{n\k }^{(1)}}v_\alpha\ket{u_{m\k }^{(0)}}_{\Omega_0} &=\sum_{n'\neq n}\sum_{s\,l}\frac{C_{sl,n'}^{(0)}(\k )C_{sl,n}^{(0)*}(\k )}{E_{n\k }^{(0)}-E_{n'\k }^{(0)}}\,e\mathcal{E} ld\bra{u_{n'\k }^{(0)}}v_\alpha\ket{u_{m\k }^{(0)}}_{\Omega_0}\\
    &=\sum_{n'\neq n}\sum_{s\,l}\frac{D_{sl,nn'}(\k )}{\Delta_{\k ,nn'}}\Tilde{\mathcal{E}}\,lv_{\alpha,n'm}(\k ),
\end{aligned}
\end{equation}
\begin{equation}
\begin{aligned}
    \bra{u_{n\k }^{(0)}}v_\alpha\ket{u_{m\k }^{(1)}}_{\Omega_0} &=\sum_{m'\neq m}\sum_{s\,l}\Big(\frac{D_{sl,mm'}(\k )}{\Delta_{\k ,mm'}}\Tilde{\mathcal{E}}\,lv_{\alpha,m'n}(\k )\Big)^*\\
    &=\sum_{m'\neq m}\sum_{s\,l}\frac{D_{sl,m'm}(\k )}{\Delta_{\k ,mm'}}\Tilde{\mathcal{E}}\,lv_{\alpha,nm'}(\k ).
\end{aligned}
\end{equation}
So, Eq.(\ref{unk0_Pumk1}) becomes:
\begin{equation}
\label{unk0_Pumk1_nota}
\begin{aligned}
    \bra{u_{n\k }^{(0)}}\partial_{k_\alpha}u_{m\k }^{(1)}\rangle_{\Omega_0} =& -\frac{v_{\alpha,nm}(\k )}{\Delta_{\k ,mn}^2}\sum_{s\,l}\big(D_{sl,mm}(\k )-D_{sl,nn}(\k )\big)\Tilde{\mathcal{E}}\,l\\
    &+\sum_{m'\neq m,n}\frac{v_{\alpha,m'm}(\k )}{\Delta_{\k ,mm'}}\sum_{s\,l}\frac{D_{sl,nm'}(\k )}{\Delta_{\k ,m'n}}\,\Tilde{\mathcal{E}}\,l\\
    &+\sum_{n'\neq n}\frac{\sum_{s\,l}D_{sl,nn'}(\k )v_{\alpha,n'm}(\k )}{\Delta_{\k ,mn}\Delta_{\k ,nn'}}\,\Tilde{\mathcal{E}}\,l+\sum_{m'\neq m}\frac{\sum_{s\,l}D_{sl,m'm}(\k )v_{\alpha,nm'}(\k )}{\Delta_{\k ,mn}\Delta_{\k ,mm'}}\,\Tilde{\mathcal{E}}\,l.
\end{aligned}
\end{equation}
Eq.(\ref{unk1_Pumk0}) becomes:
\begin{equation}
\label{unk1_Pumk0_nota}
    \bra{u_{n\k }^{(1)}}\partial_{k_\alpha}u_{m\k }^{(0)}\rangle_{\Omega_0} = \sum_{n'\neq n,m;s\,l}\frac{D_{sl,nn'}(\k )v_{\alpha,n'm}(\k )}{\Delta_{\k ,nn'}\Delta_{\k ,mn'}}\,\Tilde{\mathcal{E}}\,l.
\end{equation}
Eq.(\ref{unk0_Pumk0}) becomes:
\begin{equation}
    \bra{u_{n\k }^{(0)}}\partial_{k_\alpha}u_{m\k }^{(0)}\rangle_{\Omega_0} = \frac{v_{\alpha,nm}(\k )}{\Delta_{\k ,mn}}.
\end{equation}
In Eq.~\eqref{eq:pert_C_sln}, the first-order term becomes:
\begin{equation}
    C_{sl,n}^{(1)}(\k ) = \sum_{n'\neq n;s'l'}\frac{D_{s'l',n'n}(\k )}{\Delta_{\k ,nn'}}\,\Tilde{\mathcal{E}}\,l'C_{sl,n'}^{(0)}(\k ).
\end{equation}
So,$$
\left\{
\begin{aligned}
    C_{sl,n}^{(1)}(\k )C_{sl,n'}^{(0)*}(\k )&=\sum_{\Tilde{n}\neq n;s'l'}\frac{D_{s'l',\Tilde{n}n}(\k )}{\Delta_{\k ,n\Tilde{n}}}\,\Tilde{\mathcal{E}}\,l'D_{sl,n'\Tilde{n}}(\k ),\\
    C_{sl,n}^{(0)}(\k )C_{sl,n'}^{(1)*}(\k )&=\sum_{\Tilde{n}'\neq n';s'l'}\frac{D_{s'l',n'\Tilde{n}'}(\k )}{\Delta_{\k ,n'\Tilde{n}'}}\,\Tilde{\mathcal{E}}\,l'D_{sl,\Tilde{n}'n}(\k ),\\
    E_{n\k }^{(1)} &= \sum_{s\,l}D_{sl,nn}(\k )\,\Tilde{\mathcal{E}} l.
\end{aligned}
\right.
$$
Now we are ready to explicit Eq.~\eqref{eq:typical_term_star}:
\begin{equation*}
\begin{aligned}
    (*)&=E_{m\k }^{(0)}D_{sl,n'n}(\k )\frac{v_{\alpha,nm}(\k )}{\Delta_{\k ,mn}}\frac{v_{\beta,mn'}(\k )}{\Delta_{\k ,mn'}}\qquad\qquad[0\,{\rm th\; order}\equiv (B_1)]\\
    &+E_{m\k }^{(0)}\sum_{s'l'}\Big(\sum_{\Tilde{n}\neq n}\frac{D_{s'l',\Tilde{n}n}(\k )D_{sl,n'\Tilde{n}}(\k )}{\Delta_{\k ,n\Tilde{n}}}+\sum_{\Tilde{n}'\neq n'}\frac{D_{s'l',n'\Tilde{n}'}(\k )D_{sl,\Tilde{n}'n}(\k )}{\Delta_{\k ,n'\Tilde{n}'}}\Big)\frac{v_{\alpha,nm}(\k )v_{\beta,mn'}(\k )}{\Delta_{\k ,mn}\Delta_{\k ,mn'}}\,\Tilde{\mathcal{E}}\,l'\\&\ \ \qquad\qquad\qquad\qquad\qquad\qquad\qquad\qquad\qquad\qquad\qquad\qquad[1\,{\rm st\;order\;in}\;C_{sl,n}(\k )\equiv (B_2)]\\
    &+\sum_{s'l'}D_{s'l',mm}(\k )D_{sl,n'n}(\k )\frac{v_{\alpha,nm}(\k )v_{\beta,mn'}(\k )}{\Delta_{\k ,mn}\Delta_{\k ,mn'}}\,\Tilde{\mathcal{E}}\,l'\qquad\qquad[1\,{\rm st\;order\;in}\;E_{m\k }\equiv (B_3)]\\
    &+E_{m\k }^{(0)}D_{sl,n'n}(\k )\big(\bra{u_{n\k }^{(1)}}\partial_{k_\alpha}u_{m\k }^{(0)}\rangle_{\Omega_0}+\bra{u_{n\k }^{(0)}}\partial_{k_\alpha}u_{m\k }^{(1)}\rangle_{\Omega_0}\big)\frac{v_{\beta,mn'}(\k )}{\Delta_{\k ,mn'}}\\
    &+E_{m\k }^{(0)}D_{sl,n'n}(\k )\frac{v_{\alpha,nm}(\k )}{\Delta_{\k ,mn}}\big(\bra{u_{n'\k }^{(1)}}\partial_{k_\beta}u_{m\k }^{(0)}\rangle_{\Omega_0}+\bra{u_{n'\k }^{(0)}}\partial_{k_\beta}u_{m\k }^{(1)}\rangle_{\Omega_0}\big)^*\\
    &\qquad\qquad\qquad\qquad\qquad\qquad\qquad\qquad\qquad\qquad\qquad\qquad\qquad[1\,{\rm st\;order\;in}\;\ket{u_{n\k }}\equiv (B_4)].
\end{aligned}
\end{equation*}
So, we decompose Eq.~\eqref{eq:typical_term_star} into four contributions: $(B_1)$ to $(B_4)$. 

We now explicit the first-order correction in $\ket{u_{n\k }}$: Eq.(\ref{unk0_Pumk1_nota}) and Eq.(\ref{unk1_Pumk0_nota})
\begin{equation}
\begin{aligned}
    &\bra{u_{n\k }^{(1)}}\partial_{k_\alpha}u_{m\k }^{(0)}\rangle+\bra{u_{n\k }^{(0)}}\partial_{k_\alpha}u_{m\k }^{(1)}\rangle\equiv -iA_{\alpha,nm}^{(1)}\qquad\qquad\quad({\rm correction\;in\;Berry\;connection})\\
    &=\sum_{s'\,l'}\sum_{\Tilde{n}\neq n,m}\frac{D_{s'l',n\Tilde{n}}(\k )v_{\alpha,\Tilde{n}m}(\k )}{\Delta_{\k ,n\Tilde{n}}\Delta_{\k ,m\Tilde{n}}}\,\Tilde{\mathcal{E}}\,l'-\frac{v_{\alpha,nm}(\k )}{\Delta_{\k ,mn}^2}\sum_{s'\,l'}\big(D_{s'l',mm}(\k )-D_{s'l',nn}(\k )\big)\,\Tilde{\mathcal{E}}\,l'\\
    &+\sum_{s'l'}\sum_{\Tilde{n}\neq n,m}\frac{D_{s'l',n\Tilde{n}}(\k )v_{\alpha,\Tilde{n}m}(\k )}{\Delta_{\k ,m\Tilde{n}}\Delta_{\k ,\Tilde{n}n}}\,\Tilde{\mathcal{E}}\,l'+\sum_{s'l'}\sum_{\Tilde{n}\neq n}\frac{D_{s'l',n\Tilde{n}}(\k )v_{\alpha,\Tilde{n}m}(\k )}{\Delta_{\k ,mn}\Delta_{\k ,n\Tilde{n}}}\,\Tilde{\mathcal{E}}\,l'\\
    &+\sum_{s'l'}\sum_{\Tilde{n}\neq m}\frac{D_{s'l',\Tilde{n}m}(\k )v_{\alpha,n\Tilde{n}}(\k )}{\Delta_{\k ,mn}\Delta_{\k ,m\Tilde{n}}}\,\Tilde{\mathcal{E}}\,l'\\
    &=\sum_{s'\,l'}\Big[\sum_{\Tilde{n}\neq n}\frac{D_{s'l',n\Tilde{n}}(\k )v_{\alpha,\Tilde{n}m}(\k )}{\Delta_{\k ,mn}\Delta_{\k ,n\Tilde{n}}}+\sum_{\Tilde{n}\neq m}\frac{D_{s'l',\Tilde{n}m}(\k )v_{\alpha,n\Tilde{n}}(\k )}{\Delta_{\k ,mn}\Delta_{\k ,m\Tilde{n}}}\\&-\frac{(D_{s'l',mm}(\k )-D_{s'l',nn}(\k ))v_{\alpha,nm}(\k )}{\Delta_{\k ,mn}^2}\Big]\,\Tilde{\mathcal{E}}\,l'.
\end{aligned}
\end{equation}
Then, the first-order correction in $\ket{u_{n\k }}$ becomes
\begin{equation*}
    E_{m\k }^{(0)}D_{sl,n'n}(\k )\big[\frac{v_{\beta,mn'}(\k )}{\Delta_{\k ,mn'}}(-iA_{\alpha,nm}^{(1)})+\frac{v_{\alpha,nm}(\k )}{\Delta_{\k ,mn}(\k )}(iA_{\beta,n'm}^{(1)*})\big].
\end{equation*}

Now, we are ready for Eq.~\eqref{eq:Mzl_def_k} of the perturbed system. The first part reads
\begin{equation*}
    \frac{e}{2\hbar}{\rm Im}\{\epsilon_{z\alpha\beta}\sum_\k \sum_{s\,l}\sum_{n,n'\in{\rm occ.}}\sum_{m\in{\rm emp.}}(B_1)+(B_2)+(B_3)+(B_4)\},
\end{equation*}
the second part reads:
\begin{equation*}
    -\frac{e}{2\hbar}{\rm Im}\{\epsilon_{z\alpha\beta}\sum_\k \sum_{s\,l}\sum_{n\in{\rm occ.}}\sum_{m,m'\in{\rm emp.}}\sum_{i = 1}^4(B_i)\;{\rm with\;replacement\;}\left\{
    \begin{aligned}
        (n,n')&\rightarrow(m,m')\\m&\rightarrow n
    \end{aligned}
    \right.\}.
\end{equation*}
To the first-order in $\alpha_{zz}^{(l)}$, we only need to consider terms related to $(B_2)$, $(B_3)$ and $(B_4)$. 

Until now, the correction of the chemical potential $\mu$ has not been considered yet, which can be written as $\mu = \mu^{(0)}+\mu^{(1)}$ when field $\mathcal{E}$ is applied. Unfortunately, the correction in $\mu$ cannot be expressed solely by the information extracted from $H_0$ with PBC. This difficulty is deeply inherited from the definition of chemical potential in the grand canonical ensemble, which requires a boundary between the system and the exterior bath. For trivial band insulator, the correction in chemical potential is unimportant as long as it is still in the gap. However, for Chern insulators having conducting edge states (as in our case), the correction in chemical potential should be seriously taken into account. This reflects the importance of edge states while evaluating measurable quantities in topological insulators.

Let us first look what to change if considering the correction in $\mu$. First, only substitutions $E_{m\k }^{(0)}\rightarrow E_{m\k }^{(0)}-\mu^{(0)}$ should be made in $(B_2)$ and $(B_4)$ because the rest of terms is already of first-order. As for $(B_3)$, we should take $\mu^{(1)}$ into consideration. Our solution to $\mu^{(1)}$ is to use an auxillary system with OBC, conjugate to the system with PBC. Here, we mean ``conjugate'' by that both systems have the same level of discretization, in other words the back and forth Fourier transformation is normalized to unity. For example, if we consider a system with PBC of $\k$ mesh $100\times 100$ in the Brillouin zone, its conjugate OBC system consists of $100\times 100$ unit cells. For isolated system, a weak external electric field cannot change the particle number of system. So, if the system is half-filled, one determine $\mu^{(1)}$ by 
\begin{equation}
    \mu^{(1)} = \frac{1}{2} (E^{(1)}_{{\rm OBC},v} + E^{(1)}_{{\rm OBC},c})
\end{equation}
where $E^{(1)}_{{\rm OBC},v/c}$ is the first-order correction in the energy of the highest filled state and the lowest empty state, respectively, for the conjugate system with OBC.  
We can of course apply perturbed theory to the auxillary system, then $\mu^{(1)}$ can be formally written as
\begin{equation}
    \mu^{(1)} = \frac{1}{2} \sum_{\RR,sl} (\Tilde{C}_{sl,v}^{(0) \, *}(\RR )\Tilde{C}_{sl,v}^{(0)}(\RR ) + \Tilde{C}_{sl,c}^{(0) \, *}(\RR )\Tilde{C}_{sl,c}^{(0)}(\RR ) )\,\Tilde{\mathcal{E}} l
    % \frac{1}{2}\sum_{s'l'}(D_{s'l',N^2N_{\L }\,N^2N_{\L }}+D_{s'l',N^2N_{\L }+1\,N^2N_{\L }+1})\,\Tilde{\mathcal{E}}\,l'
\end{equation}
where $\Tilde{C}_{sl,v/c}^{(0)}(\RR)$ is the form factor $\braket{sl,\RR}{\psi_{{\rm OBC},v/c}^{(0)}}$ for the conjugate system. This expression singles out layer index, which would be useful in the next section. Meanwhile, we can also factorize the layer index out for $(B_3)$  
% where $N_{\L }$ is the total layer number of the system, and there are $N$ unit cells in a monolayer. So 
\begin{equation}
    (B_3) = \sum_{s'l'}\left[D_{s'l',mm}(\k )-\frac{\sum_\RR \Tilde{C}_{s'l',v}^{(0) \, *}(\RR )\Tilde{C}_{s'l',v}^{(0)}(\RR ) + \Tilde{C}_{s'l',c}^{(0) \, *}(\RR )\Tilde{C}_{s'l',c}^{(0)}(\RR )}{2}\right] D_{sl,n'n}(\k )\frac{v_{\alpha,nm}(\k )v_{\beta,mn'}(\k )}{\Delta_{\k ,mn}\Delta_{\k ,mn'}}\,\Tilde{\mathcal{E}}\,l'.
\end{equation}
Then, the formula for $\alpha_{zz}^{(l)}$ is derived by singling out the electric field $\mathcal{E}$ and breaking the sum over layer index into parts for each layer, respectively, in Eq.~\eqref{eq:Mzl_def_k}. 

Note that the derivation above requires energy levels to be non-degenerate, otherwise the denominator would be vanishing. This should not worry us since the energy bands are non-degenerate except at some high-symmetry points if interlayer coupling is turned on. We observe that the energy bands are all non-degenerate for AA stacking and degenerate at $K$ for AB stacking. The degeneracy at $K$ is protected by space symmetry of the slab. In principle, we can choose a gauge such that the Hamiltonian is block diagonal at this degenerate point then separately apply non-degenerate perturbation theory as above. In practice, we omit the contribution from the degeneracy point which would affect too much the accuracy as long as the mesh is sufficiently fine.

\section{Proof of layer index gauge invariance}
We see from above that layer index appears explicitly in the expressions from $(B_1)$ to $(B_4)$. It seems that the results would depend on the ``gauge choice'' of layer index, which is a priori arbitrary. In other words, if there are five layers, we can name them as $1,2,3,4,5$ or $-2,-1,0,1,2$ or in other similar ways. In this section, we will prove that the results of $\alpha_{zz}^{(l)}$ are independent of the gauge choice of layer index. Without loss of generality, this is equivalent to prove that the expression of $(B_2)$, $(B_3)$ and $(B_4)$ is vanishing if setting $l'=1$. Let us see what these terms become. For $(B_2)$, we have 
\begin{equation}
\begin{aligned}
    (B_2) &= (E_{m\k }^{(0)}-\mu^{(0)})\Big(\sum_{\Tilde{n}\neq n}\frac{\sum_{s'l'}D_{s'l',\Tilde{n}n}(\k )D_{sl,n'\Tilde{n}}(\k )}{\Delta_{\k ,n\Tilde{n}}}+\sum_{\Tilde{n}'\neq n'}\frac{\sum_{s'l'}D_{s'l',n'\Tilde{n}'}(\k )D_{sl,\Tilde{n}'n}(\k )}{\Delta_{\k ,n'\Tilde{n}'}}\Big)\\&\times\frac{v_{\alpha,nm}(\k )v_{\beta,mn'}(\k )}{\Delta_{\k ,mn}\Delta_{\k ,mn'}}\,\Tilde{\mathcal{E}}\\
    &=(E_{m\k }^{(0)}-\mu^{(0)})\Big(\sum_{\Tilde{n}\neq n}\frac{\delta_{\Tilde{n}n}(\k )D_{sl,n'\Tilde{n}}(\k )}{\Delta_{\k ,n\Tilde{n}}}+\sum_{\Tilde{n}'\neq n'}\frac{\delta_{n'\Tilde{n}'}D_{sl,\Tilde{n}'n}(\k )}{\Delta_{\k ,n'\Tilde{n}'}}\Big)\\&\times\frac{v_{\alpha,nm}(\k )v_{\beta,mn'}(\k )}{\Delta_{\k ,mn}\Delta_{\k ,mn'}}\,\Tilde{\mathcal{E}} = 0.
\end{aligned}
\end{equation}
As for $(B_4)$,
\begin{equation}
\begin{aligned}
    -iA_{\alpha,nm}^{(1)\,*}&=\Big[\sum_{\Tilde{n}\neq n}\frac{\sum_{s'\,l'}D_{s'l',n\Tilde{n}}(\k )v_{\alpha,\Tilde{n}m}(\k )}{\Delta_{\k ,mn}\Delta_{\k ,n\Tilde{n}}}+\sum_{\Tilde{n}\neq m}\frac{\sum_{s'\,l'}D_{s'l',\Tilde{n}m}(\k )v_{\alpha,n\Tilde{n}}(\k )}{\Delta_{\k ,mn}\Delta_{\k ,m\Tilde{n}}}\\&-\frac{(\sum_{s'\,l'}D_{s'l',mm}(\k )-\sum_{s'\,l'}D_{s'l',nn}(\k ))v_{\alpha,nm}(\k )}{\Delta_{\k ,mn}^2}\Big]\,\Tilde{\mathcal{E}}\\
    &=\Big[\sum_{\Tilde{n}\neq n}\frac{\delta_{n\Tilde{n}}(\k )v_{\alpha,\Tilde{n}m}(\k )}{\Delta_{\k ,mn}\Delta_{\k ,n\Tilde{n}}}+\sum_{\Tilde{n}\neq m}\frac{\delta_{\Tilde{n}m}(\k )v_{\alpha,n\Tilde{n}}(\k )}{\Delta_{\k ,mn}\Delta_{\k ,m\Tilde{n}}}\\&-\frac{(\delta_{mm}-\delta_{nn})v_{\alpha,nm}(\k )}{\Delta_{\k ,mn}^2}\Big]\,\Tilde{\mathcal{E}} = 0,
\end{aligned}
\end{equation}
thus, $(B_4) = 0$. And
\begin{equation}
\begin{aligned}
    (B_3) &= \big[\sum_{s'l'}D_{s'l',mm}(\k )-\frac{\sum_{s'l'\RR} \Tilde{C}_{s'l',v}^{(0) \, *}(\RR )\Tilde{C}_{s'l',v}^{(0)}(\RR ) + \Tilde{C}_{s'l',c}^{(0) \, *}(\RR )\Tilde{C}_{s'l',c}^{(0)}(\RR )}{2}\big]\\&\times D_{sl,n'n}(\k )\frac{v_{\alpha,nm}(\k )v_{\beta,mn'}(\k )}{\Delta_{\k ,mn}\Delta_{\k ,mn'}}\,\Tilde{\mathcal{E}}\\
    &=\big[\delta_{mm}-1\big]\,D_{sl,n'n}(\k )\frac{v_{\alpha,nm}(\k )v_{\beta,mn'}(\k )}{\Delta_{\k ,mn}\Delta_{\k ,mn'}}\,\Tilde{\mathcal{E}}\\
    &=0.
\end{aligned}
\end{equation}
So, the gauge invariance of layer index for $\alpha_{zz}^{(l)}$ is proved.

\section{Proof of the quantization rule in the interlayer decoupled case}
In this section, we will prove rigorously the quantization rule Eq.~(7) in the main text for the interlayer decoupled case, where the total Hamiltonian is obviously block-diagonal in the layer space. Each diagonal block is the Hamiltonian of every monolayer and they are all the same. If we diagonalize the full Hamiltonian, there are exactly $N_{\L}$ degenerate energy valence bands and $N_{\L}$ degenerate energy conduction bands if the slab consists of $N_\L$ layers. Without loss of generality, we can associate valence band $n=1,\dots,N_\L$ with layer index $l=1 \dots N_{\L}$, respectively. Similarly, we can also do the same for the conduction band $n=N_\L+1,\dots,2N_\L$ with layer index $l=1 \dots N_{\L}$, respectively. This leads to
\begin{equation}
    \label{eq:normalization_decoupled}
\begin{aligned}
        \sum_{s}D_{sl,n'n} = \sum_{s}C_{sl,n'}^*C_{sl,n} = \delta_{nn'}\delta_{l,n/n-N_\L}
\end{aligned}
\end{equation}
for the band $n$ is the valence/conduction band, respectively.

Thus,
\begin{equation}
\begin{aligned}
    \sum_{s}(B_2) &= E_{m\k }^{(0)}\sum_{l'}\Big(\sum_{\Tilde{n}\neq n}\frac{\sum_{s'}D_{s'l',\Tilde{n}n}(\k )\sum_{s}D_{sl,n'\Tilde{n}}(\k )}{\Delta_{\k ,n\Tilde{n}}}+\sum_{\Tilde{n}'\neq n'}\frac{\sum_{s'}D_{s'l',n'\Tilde{n}'}(\k )\sum_{s}D_{sl,\Tilde{n}'n}(\k )}{\Delta_{\k ,n'\Tilde{n}'}}\Big)\\
    &\times\frac{v_{\alpha,nm}(\k )v_{\beta,mn'}(\k )}{\Delta_{\k ,mn}\Delta_{\k ,mn'}}\,\Tilde{\mathcal{E}}\,l'\\
    &=E_{m\k }^{(0)}\sum_{l'}\Big(\sum_{\Tilde{n}\neq n}\frac{\delta_{n\Tilde{n}}\,\delta_{l',n}\,\delta_{n'\Tilde{n}}\,\delta_{l,n'}}{\Delta_{\k ,n\Tilde{n}}}+\sum_{\Tilde{n}'\neq n'}\frac{\delta_{\Tilde{n}'n'}\,\delta_{l',n'}\,\delta_{\Tilde{n}'n}\,\delta_{l,n}}{\Delta_{\k ,n'\Tilde{n}'}}\Big)\\
    &\times\frac{v_{\alpha,nm}(\k )v_{\beta,mn'}(\k )}{\Delta_{\k ,mn}\Delta_{\k ,mn'}}\,\Tilde{\mathcal{E}}\,l'\\
    &=0
\end{aligned}
\end{equation}

Meanwhile,
\begin{equation*}
\begin{aligned}
    \sum_{s}(B_4) &= E_{m\k }^{(0)}\sum_{s}D_{sl,n'n}(\k )\big[(-iA_{\alpha,nm}^{(1)})\frac{v_{\beta,mn'}(\k )}{\Delta_{\k ,mn'}}+\frac{v_{\alpha,nm}(\k )}{\Delta_{\k ,mn}(\k )}(iA_{\beta,n'm}^{(1)*})\big]\\
    &= E_{m\k }^{(0)}\,\delta_{n'n}\,\delta_{l,n}\big[(-iA_{\alpha,nm}^{(1)})\frac{v_{\beta,mn'}(\k )}{\Delta_{\k ,mn'}}+\frac{v_{\alpha,nm}(\k )}{\Delta_{\k ,mn}(\k )}(iA_{\beta,n'm}^{(1)*})\big],
\end{aligned}
\end{equation*}
\begin{equation*}
\begin{aligned}
    -iA_{\alpha,nm}^{(1)}&=\sum_{l'}\Big[\sum_{\Tilde{n}\neq n}\frac{\sum_{s'}D_{s'l',n\Tilde{n}}(\k )v_{\alpha,\Tilde{n}m}(\k )}{\Delta_{\k ,mn}\Delta_{\k ,n\Tilde{n}}}+\sum_{\Tilde{n}\neq m}\frac{\sum_{s'}D_{s'l',\Tilde{n}m}(\k )v_{\alpha,n\Tilde{n}}(\k )}{\Delta_{\k ,mn}\Delta_{\k ,m\Tilde{n}}}\\
    &\quad -\frac{\big(\sum_{s'}D_{s'l',mm}(\k )-\sum_{s'}D_{s'l',nn}(\k )\big)v_{\alpha,nm}(\k )}{\Delta_{\k ,mn}^2}\Big]\,\Tilde{\mathcal{E}}\,l'\\
    &=\sum_{l'}\Big[\sum_{\Tilde{n}\neq n}\frac{\delta_{n\Tilde{n}}\,\delta_{l',n}v_{\alpha,\Tilde{n}m}(\k )}{\Delta_{\k ,mn}\Delta_{\k ,n\Tilde{n}}}+\sum_{\Tilde{n}\neq m}\frac{\delta_{\Tilde{n}m}\,\delta_{l',m-N_\L}v_{\alpha,n\Tilde{n}}(\k )}{\Delta_{\k ,mn}\Delta_{\k ,m\Tilde{n}}}-\frac{\big(\delta_{l',m-N_\L}-\delta_{l',n}\big)v_{\alpha,nm}(\k )}{\Delta_{\k ,mn}^2}\Big]\,\Tilde{\mathcal{E}}\,l'\\
    & = -\frac{(m-N_\L-n)v_{\alpha,nm}(\k )}{\Delta_{\k ,mn}^2}\,\Tilde{\mathcal{E}}.
\end{aligned}
\end{equation*}
Thus,
\begin{equation}
\label{B4_contribution}
\begin{aligned}
    \sum_{s}\sum_{n,n'\in{\rm occ.}}\sum_{m\in{\rm emp}}(B_4)&=\sum_{n\in{\rm occ.}}\sum_{m\in{\rm emp}}E_{m\k }^{(0)}\,\delta_{l,n}\big[(-iA_{\alpha,nm}^{(1)})\frac{v_{\beta,mn}(\k )}{\Delta_{\k ,mn}}+\frac{v_{\alpha,nm}(\k )}{\Delta_{\k ,mn}(\k )}(iA_{\beta,nm}^{(1)*})\big]\\
    &=-2\sum_{n\in{\rm occ.}}\sum_{m\in{\rm emp}}E_{m\k }^{(0)}\,\delta_{l,n}(m-n-N_\L)\frac{v_{\alpha,nm}(\k )v_{\beta,mn}(\k )}{\Delta_{\k ,mn}^3}\,\Tilde{\mathcal{E}}.
\end{aligned}
\end{equation}
If $m-N_\L = n$, Eq.~(\ref{B4_contribution}) equals $0$ obviously. If $m-N_\L\neq n$, since $v_{\alpha,nm} = \bra{u_{n\k }^{(0)}}(\partial_{k_\alpha}H_\k )\ket{u_{m\k }^{(0)}}_{\Omega_0}$ is also block-diagonal, the expression must be zeros as well. To sum up, $(B_4)$ also contributes $0$. 

Now we turn to $(B_3)$:
\begin{equation*}
\begin{aligned}
    \sum_{s}\sum_{n,n'\in{\rm occ.}}\sum_{m\in {\rm emp}}(B_3) &=\sum_{n,n'\in{\rm occ.}}\sum_{m\in {\rm emp}}\sum_{l'}\big[\sum_{s'}D_{s'l',mm}(\k )-\frac{\sum_{s'\RR} \Tilde{C}_{s'l',v}^{(0) \, *}(\RR )\Tilde{C}_{s'l',v}^{(0)}(\RR ) + \Tilde{C}_{s'l',c}^{(0) \, *}(\RR )\Tilde{C}_{s'l',c}^{(0)}(\RR )}{2}\big]\\
    &\quad \times\sum_{s}D_{sl,n'n}(\k )\frac{v_{\alpha,nm}(\k )v_{\beta,mn'}(\k )}{\Delta_{\k ,mn}\Delta_{\k ,mn'}}\,\Tilde{\mathcal{E}}\,l'\\
    &=\sum_{n,n'\in{\rm occ.}}\sum_{m\in {\rm emp}}\sum_{l'}\big[\delta_{l',m-N_\L}-\frac{\delta_{l',N_{\L }}+\delta_{l',1}}{2}\big]\delta_{n'n}\delta_{l,n}\,\frac{v_{\alpha,nm}(\k )v_{\beta,mn'}(\k )}{\Delta_{\k ,mn}\Delta_{\k ,mn'}}\,\Tilde{\mathcal{E}}\,l'\\
    &=\sum_{n\in{\rm occ.}}\sum_{m\in{\rm emp}}\big[m-N_\L-\frac{N_\L + 1}{2}\big]\delta_{l,n}\frac{v_{\alpha,nm}(\k )v_{\beta,mn}(\k )}{\Delta_{\k ,mn}^2}\,\Tilde{\mathcal{E}}\\
    &=\sum_{m\in{\rm emp}} \delta_{m,l+N_\L} \big[l-\frac{N_\L + 1}{2}\big] \frac{v_{\alpha,lm}(\k )v_{\beta,ml}(\k )}{\Delta_{\k ,ml}^2}\,\Tilde{\mathcal{E}},
\end{aligned}
\end{equation*}
where we use a similar mapping as Eq.~\eqref{eq:normalization_decoupled} for the system with OBC to derive the first equality. The last equality comes from the fact that $v_{\alpha,nm} = \bra{u_{n\k }^{(0)}}(\partial_{k_\alpha}H_\k )\ket{u_{m\k }^{(0)}}_{\Omega_0}$ is also block-diagonal. Note that this expression is precisely the Chern number of monolayer. Therefore, if there are odd layers in system, the response is integer quantized; if $N_\L$ is even, the respectively is then quantized to half-integer. Then, the quantization rule Eq.~(7) in the main text is automatically proved for the interlayer decoupled case.

\section{Details on numerical calculations and discussions on finite-size effect}
In this section, we provide more details on the numerical calculations of $\alpha_{zz}^{(l)}$ and layer Chern number. We also discuss finite-size effect on the improvement of quantization.

\subsection{Layer-resolved orbital magnetoelectric response of slab}
In a slab system, we calculate $\alpha_{zz}^{(l)}$ using $\partial M_z^{(l)} / \partial \mathcal{E}$. We employ three approaches in our study. The used mathematical expressions are all given above. 

The first approach is that we first calculate $M_z^{(l)}$ in the absence and in the presence of weak electric field using Eq.~\eqref{eq:Mzl_def_brute}. Then, we get $\alpha_{zz}^{(l)}$ by finite difference. In practice, we calculate $M_z^{(l)}$ for a series of weak electric field such that the response is still in the linear regime. Then, we fit the results to get the slope of the straight line and thus $\alpha_{zz}^{(l)}$. This approach is uniquely suitable for a slab with OBC in the $x,y$ plane. We use this approach to test the robustness of the quantization rule against disorder. 

The second approach is similar to the first one except using Eq.~\eqref{eq:Mzl_def_k}, then get $\alpha_{zz}^{(l)}$ by finite difference. We use this for a slab with PBC in the $x,y$ plane. The obtained $\alpha_{zz}^{(l)}$ is quantized to integers if $N_{\rm L}$ odd or to half-integers if $N_{\rm L}$ even, as shown in Table~\ref{tab:layer_resolved_PBC}. The distribution is symmetric to the central plane of the system and the total $\alpha_{zz}$ is indeed vanishing as dictated by mirror symmetry.

\begin{table}%[H] add [H] placement to break table across pages
    \caption{Layer-resolved $\alpha_{zz}^{(l)}$ in units of $e^2/h$ obtained by numerical finite difference and those evaluated by perturbation theory given in the parenthesis. We consider $AA$,$AB$ stackings and $N_{\rm{L}} = 5,6$.}
    \label{tab:layer_resolved_PBC}
    \begin{ruledtabular}
    \begin{tabular}{c| c c c c c c}
    Layer Index &1&2&3&4&5&6\\
    \hline
    AA,5&1.9893 (1.9905)&0.9971 (0.9982)&-0.0010 (9.80E-14)&-0.9990 (-0.9982)&-1.9918 (-1.9905)&\diagbox[dir=NW]{}{}\\
    AB,5&1.9289 (1.9739)&0.9525 (0.9973)&-0.0456 (5.29E-05)&-1.0424 (-0.9974)&-2.0194 (-1.9739)&\diagbox[dir=NW]{}{}\\
    AA,6&2.5365 (2.4894)&1.5444 (1.4974)&0.5464 (0.4992)&-0.4519 (-0.4992)&-1.4502 (-1.4974)&-2.4423 (-2.4894)\\
    AB,6&2.4863 (2.4727)&1.5098 (1.4962)&0.5123 (0.4987)&-0.4855 (-0.4990)&-1.4828 (-1.4963)&-2.4592 (-2.4727)\\
    \end{tabular}
    \end{ruledtabular}
\end{table}

\begin{figure}[h]
    \includegraphics[width=0.95\textwidth]{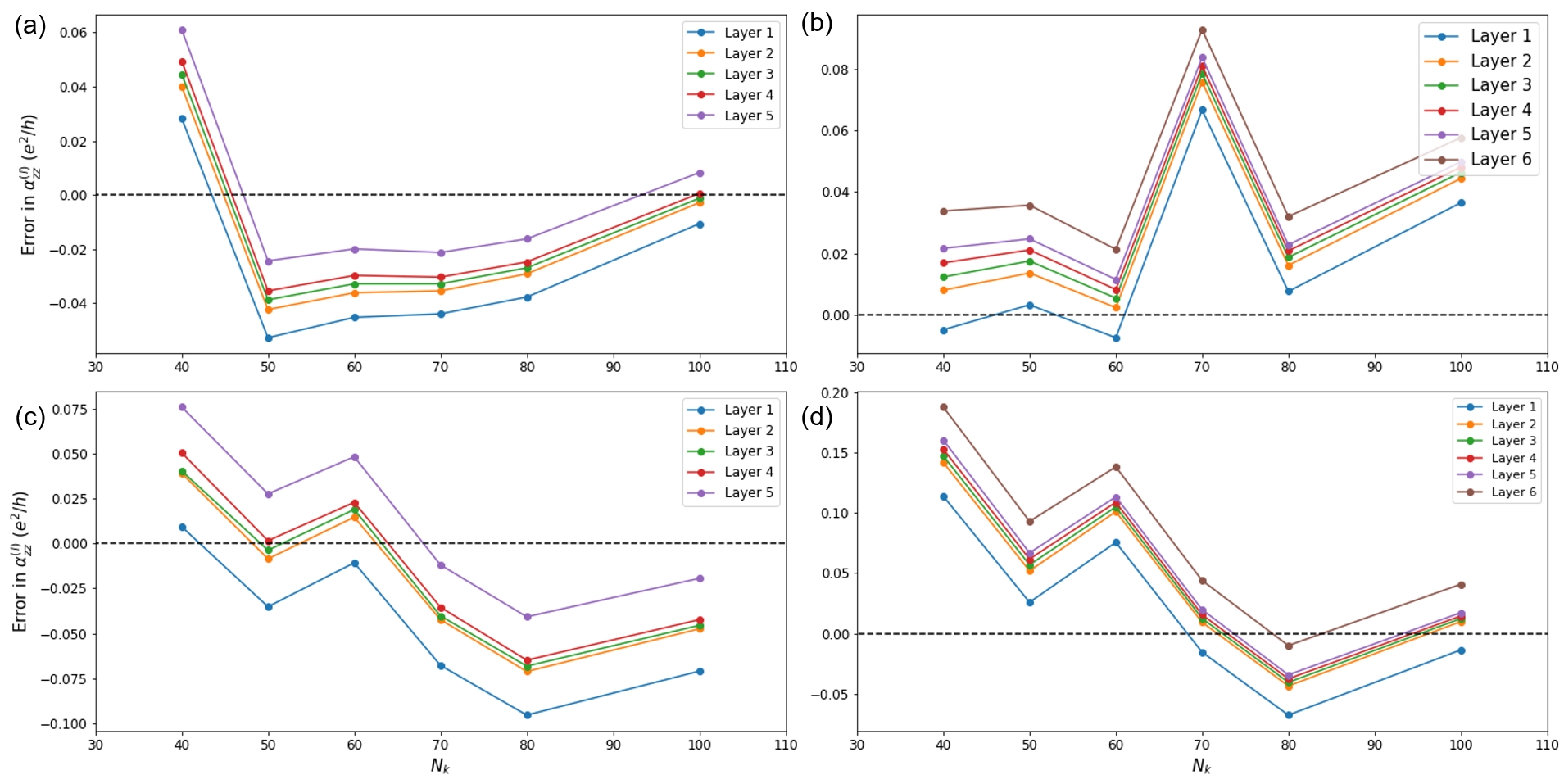}
    \caption{Error in $\alpha_{zz}^{(l)}$ with respect to the quantized value using finite-difference method. (a) AA stacking, five layers; (b) AA stacking, six layers; (c) AB stacking, five layers; (a) AB stacking, six layers.}
    \label{fig:Mzl_fd}
\end{figure} 
In Fig.~\ref{fig:Mzl_fd}, we show how the quantization is improved when we refine the $\k$ mesh. We see that the improvement is not monotonic but oscillating. The main source of error comes from the determination of chemical potential given by an auxillary system with OBC. 

Note that this is also the method we use to evaluate $\alpha_{zz}^{(l)}$ in the presence of substrate. The obtained $\alpha_{zz}^{(l)}$ are given in Table~\ref{tab:substrate} where $\k$ mesh is $90 \times 90$. We see that $\alpha_{zz}^{(l)}$ is not quantized in the presence of substrate and interlayer coupling because of inversion or mirror symmetry breaking. However, the difference between neighboring $\alpha_{zz}^{(l)}$ is well quantized as shown in Table~\IV\  of the main text.

\begin{table}
    \centering
    \caption{Layer-resolved response in units of $e^2/h$ in systems with substrate.}
    \label{tab:substrate}
    \begin{ruledtabular}
    \begin{tabular}{c| c c c c c}
    $\alpha_{zz}^{(l)}$ & $l=1$ & $l=2$ & $l=3$ & $l=4$ & $l=5$\\
    \hline
    $t_3=0,t_4=0$ &2.4930&1.4958&0.4986&-0.4986&-1.4958\\
    % \cline{2-2}\cline{5-9}
    $t_3=1/6,t_4=0$ &2.1742&1.1771&0.1799&-0.8173&-1.8145\\
    % \cline{2-2}\cline{5-9}
    $t_3=0.3,t_4=0$ &1.8602&0.8630&-0.1342&-1.1314&-2.1286\\
    % \cline{2-2}\cline{4-9}
    $t_3=1/6,t_4=3t_3/4$ &2.3786&1.3852&0.3874&-0.6104&-1.6041\\
    % \cline{2-2}\cline{5-9}
    $t_3=0.3,t_4=3t_3/4$ &1.9530&0.9690&-0.0306&-1.0296&-2.0144\\
    \end{tabular}
    \end{ruledtabular}
\end{table}

The third approach is to calculate $\alpha_{zz}^{(l)}$ directly applying perturbation formula given above. In Fig.~\ref{fig:Mzl_pert}, we show how the quantization is improved when we refine the $\k$ mesh. Compared to the second approach, one advantage of the perturbative approach is that the quantization gradually improves as the $k$ mesh becomes finer. We also note that another advantage is to enforce the mirror symmetry in the results such that $\alpha_{zz}^{(l)}$ is symmetric with respect to the central plane.
\begin{figure}[h]
    \includegraphics[width=0.95\textwidth]{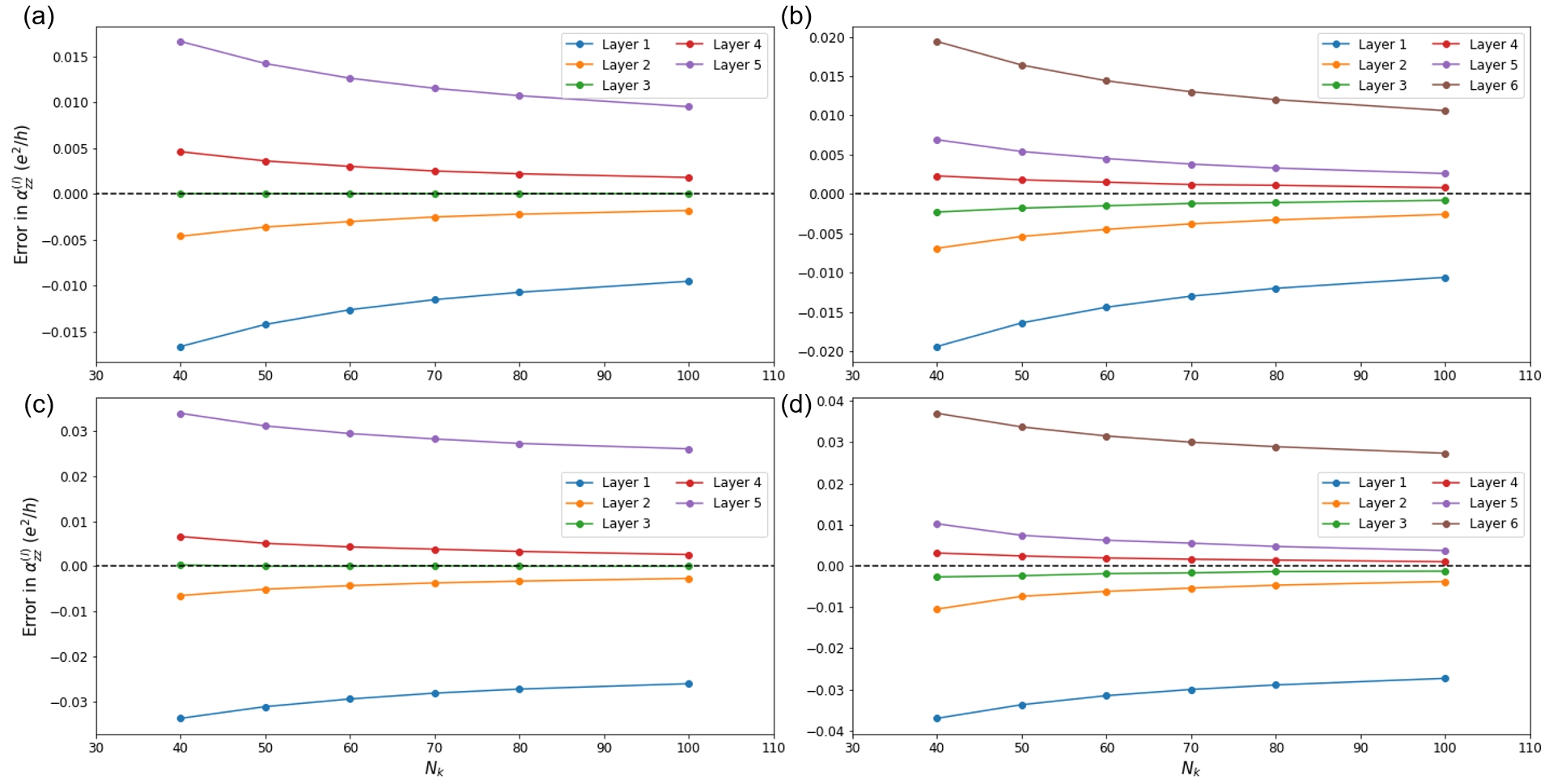}
    \caption{Error in $\alpha_{zz}^{(l)}$ with respect to the quantized value using perturbation theory. (a) AA stacking, five layers; (b) AA stacking, six layers; (c) AB stacking, five layers; (a) AB stacking, six layers.}
    \label{fig:Mzl_pert}
\end{figure}

\subsection{Layer-resolved orbital magnetoelectric response of 3D bulk}
In 3D bulk system, we calculate $\alpha_{zz}^{(l)}$ using $\partial P_z^{(l)} / \partial B$. However, the layer-resolved polarization $P_z^{(l)}$ is only well defined if the Wannier centers can be unambiguously associated with a given layer. We show in Fig.~\ref{fig:wannier_band} that this is indeed the case. We see that the Wannier centers are always very close to the position of 2D layers.
\begin{figure}
    \includegraphics[width=0.95\textwidth]{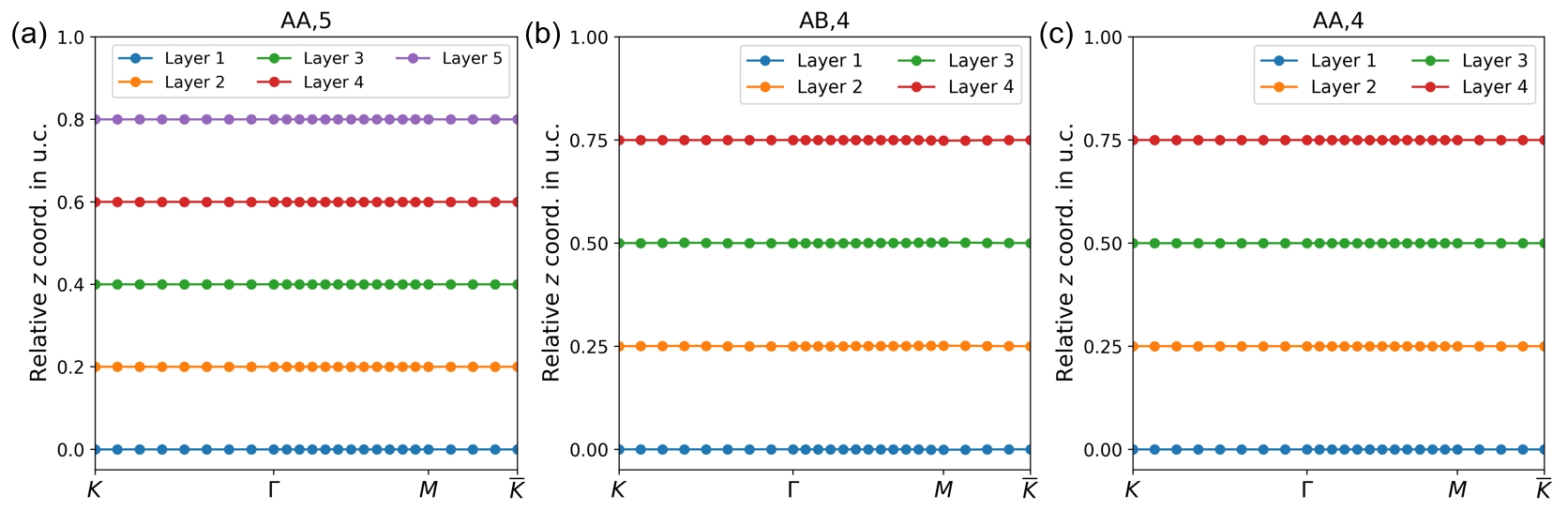}
    \caption{Band structures of Wannier centers for all the cases we show in the main text: (a) AA stacking, five layers; (b) AA stacking, four layers; (c) AB stacking, four layers.}
    \label{fig:wannier_band}
\end{figure} 

In practice, we construct a magnetic unit cell to restore magnetic translation symmetry for a magnetic field commensurate with the atomic unit cell $p/q=1/30$. To calculate $\alpha_{zz}^{(l)}$, we need to solve the Hofstadter spectrum of the Hamiltonian in the absence and in the presence of such a weak magnetic field. To be coherent in the formalism, we also use the same magnetic unit cell in the absence of magnetic field, namely we consider a magnetic field $p/q=0/30$. Then, we can calculate the layer-resolved polarization using Eq.~(11) in the main text. Then, we do a finite difference between $P_z^{(l)}$ in the absence and in presence of magnetic field and average over the 2D magnetic Brillouin zone to get the final results.

\subsection{Layer Chern number}
In the main text, we argue that the quantization rule of $\alpha_{zz}^{(l)}$ relies on quantized layer Chern number. In this subsection, we numerically check the quantization of layer Chern number using Eq.~\eqref{eq:layer_chern}. Here, the interlayer coupling is $t_3=0.2$ and $t_4=0.15$ and the intralayer parameters are still the same as in the main text. In Fig.~\ref{fig:layer_chern}, we show the error of quantization for layer Chern number is gradually improved as $\k$ mesh becomes finer. The error in layer Chern number is solely due to the finite mesh. The evaluation of layer Chern number using Eq.~\eqref{eq:layer_chern} does not reflect the topological character, therefore harder to converge compared to using the plaquette method to calculate Chern number. For example, if we set interlayer coupling to be zero, the layer Chern number is exactly the 2D Chern number should be quantized as a topological invariant. However, for a fine mesh $100\times 100$, the evaluation of Chern number using the plaquette method is quantized to precision $10^{-5}$, while the evaluation using Eq.~\eqref{eq:layer_chern} is only to precision $10^{-3}$.
\begin{figure}
    \includegraphics[width=0.95\textwidth]{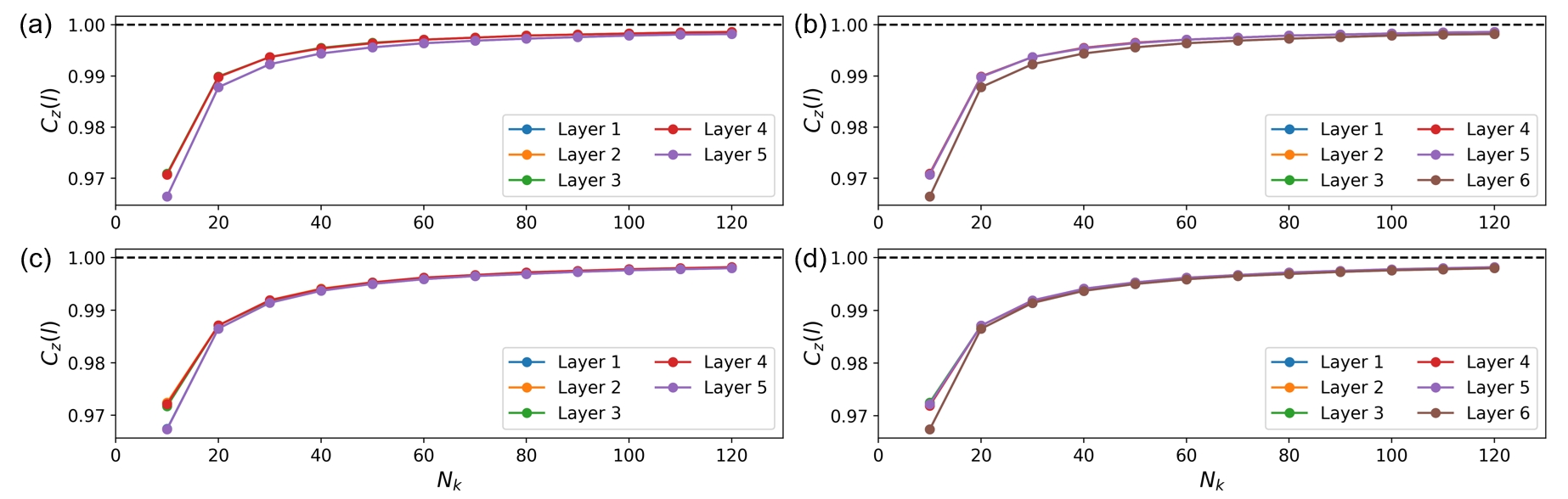}
    \caption{Error in $C_z(l)$ with respect to the quantized value. (a) AA stacking, five layers; (b) AA stacking, six layers; (c) AB stacking, five layers; (a) AB stacking, six layers.}
    \label{fig:layer_chern}
\end{figure} 

We also check that the layer Chern number is still well quantized in the presence of substrate. Here, we set the substrate on-site energy to be $t_2/2$ and $t_3=0.2$, $t_4=0.15$ as before. The $\k$ mesh is $100 \times 100$. As shown in Table, the quantization of layer Chern number is good for two stackings compared to that without substrate. As a reference, the layer Chern number without interlayer coupling turns out to be 0.9975 in the same mesh. It is remarkable that the substrate has no effect on the value of layer Chern number whose quantization is not protected by inversion or mirror symmetry.
\begin{table}%[H] add [H] placement to break table across pages
    \caption{Layer Chern number $C_{z}(l)$ with substrate and those without substrate in the parenthesis.}
    \label{tab:layer_chern_PBC}
    \begin{ruledtabular}
    \begin{tabular}{c| c c c c c c}
    Layer Index &1&2&3&4&5&6\\
    \hline
    AA,5&0.9979(0.9979)&0.9983(0.9983)&0.9983(0.9983)&0.9983(0.9983)&0.9979(0.9979)&\diagbox[dir=NW]{}{}\\
    AB,5&0.9976(0.9976)&0.9977(0.9978)&0.9978(0.9977)&0.9978(0.9978)&0.9976(0.9976)&\diagbox[dir=NW]{}{}\\
    AA,6&0.9979(0.9979)&0.9983(0.9983)&0.9983(0.9983)&0.9983(0.9983)&0.9983(0.9983)&0.9979(0.9979)\\
    AB,6&0.9976(0.9976)&0.9977(0.9978)&0.9977(0.9977)&0.9977(0.9977)&0.9978(0.9978)&0.9976(0.9976)\\
    \end{tabular}
    \end{ruledtabular}
\end{table} 

\section{Off-diagonal OME response of Chern insulators}
Let us consider a 3D Chern insulator with Chern vector $\textbf{C} = (0,0,C_z)$, where $C_z = C (C\in \mathbb{Z}, C\neq 0)$ with interlayer distance $d$. What would happen if a non-disruptive in-plane electric field is applied assuming the interlayer coupling is weak? Since 3D Chern insulator is adiabatically connected to trivial stacking of 2D Chern insulators, the response to in-plane electric field (corresponding to off-diagonal $\alpha_{xz}$ or similar components) should behave similarly as in the 2D case. Specifically, if a uniform electric field $\mathcal{E}_x$ is applied along the $x$ axis, a transverse anomalous Hall current flowing along the $y$ direction would be induced due to the non-zero Chern number, with $j_y=C e^2/(d h) \mathcal{E}_x$. The bulk anomalous Hall current would naturally generate orbital magnetization $\mathbf{M}$ via $\nabla\times \mathbf{M}=\mathbf{j}$, giving rise to an off-diagonal OME response, with its response coefficient varying linearly in real space, i.e., $\partial_x \alpha_{zx} = -Ce^2/h d$. Here $e$ is fundamental charge, $h$ is Planck constant, and $d$ is interlayer distance.

\section{First-principle DFT calculations for $\rm{MnBi}_2 \rm{Te}_4$ class with strain}

First-principle density functional theory (DFT) calculations for Mn(Bi/Sb)$_2$Te$_4$ [(Bi/Sb)$_2$Te$_3$]$_n$ thin film with $n=0,1,2$ are performed with Vienna Ab-initio Simulation Package (VASP) \cite{VASP} using the projector-augmented wave (PAW) method and the plane-wave basis with an energy cutoff of 450 eV. The Perdew-Burke-Ernzerhof (PBE) exchange-correlation functional \cite{PBE} is used in combination with GGA+$U$ method to treat the localized $d$ orbitals of Mn, with $U = 3\,\rm{eV}$ set in all calculations. The Brillouin zone is sampled by a $16 \times 16 \times 1$ $\Gamma$-centered Monkhorst-Pack $\mathbf{k}$-point mesh. Structure optimizations are performed with a force criterion of $0.01$ eV/\AA. 

The in-plane strain along the $a$ axis is applied by scaling the length of the $a$ axis, and then relaxing the atomic positions with the scaled axis fixed. The out-of-plane compression along the $c$ axis are applied as follows: we first reduce the height of the top layer atoms and then, while fixing both the top and bottom layer atoms, relax the atomic positions between these two layers.

\subsection{Chern number of $\rm{MnBi}_2 \rm{Te}_4$ class}

Once the electronic structure is converged, the Bloch states are projected to the Wannier functions of the Bi-$6p$ (or Sb-$5p$), and Te-$5p$ orbitals to generate the tight-binding Hamiltonian for identifying the Chern number \cite{wannier90_compphyscomm,wannier90_rmp}. Searching among many possibilities of Mn(Bi/Sb)$_2$Te$_4$[(Bi/Sb)$_2$Te$_3$]$_n$ class, it turns out that the best building block for the construction of 3D Chern insulator slabs is three-layer $\rm{MnBi}_2 \rm{Te}_4$ (see Table~\ref{tab:MBT}). Then, we can construct the tight-binding Hamiltonian of 3D Chern insulator slab using three-layer $\rm{MnBi}_2 \rm{Te}_4$ as repetition entity. Suppose the adjacent entity layer is separated by a spacer of thickness of single layer $\rm{MnBi}_2 \rm{Te}_4$. After classifying the Wannier centers into their corresponding sublayer in three-layer $\rm{MnBi}_2 \rm{Te}_4$ entity, we retrieve the interlayer hopping between the two outmost sublayers, served as the Hermitian conjugate of the inter-entity layer coupling.

\begin{table}
    \centering
    \caption{Chern numbers of Mn(Bi/Sb)$_2$Te$_4$[(Bi/Sb)$_2$Te$_3$]$_n$-class thin films under different strain magnitudes. ``FM" and ``AFM" stand for interlayer ferromagnetic and anti-ferromagnetic configurations, respectively.}
    \label{tab:MBT}
    \tabcolsep=3mm
    \renewcommand\arraystretch{1.5}
    \begin{tabular}{l|ccccccccc}
        \hline
        \hline
        $\mathrm{system\,(layers\mid magnetism\mid strain\,axis)}$& \multicolumn{1}{c|}{-2.0\%} & \multicolumn{1}{c|}{-1.5\%} & \multicolumn{1}{c|}{-1.0\%} & \multicolumn{1}{c|}{-0.5\%} & \multicolumn{1}{c|}{0.0\%} & \multicolumn{1}{c|}{0.5\%} & \multicolumn{1}{c|}{1.0\%} & \multicolumn{1}{c|}{1.5\%} & 2.0\% \\ \hline
        $\mathrm{\quad\quad\quad MnBi_2Te_4\,\,\,(\,2\mid \,\,\,FM\,\,\mid a\,)}$ & \multicolumn{8}{c|}{0}& 1\\ \hline
        $\mathrm{\quad\quad\quad MnBi_2Te_4\,\,\,(\,3\mid AFM\,\mid a\,)}$ & \multicolumn{8}{c|}{0}& 1\\ \hline
        $\mathrm{\quad\quad\quad MnSb_4Te_7\,\,\,(\,2\mid \,\,\,FM\,\,\mid a\,)}$ & \multicolumn{1}{c|}{1}& \multicolumn{8}{c}{0}\\ \hline
        $\mathrm{\quad\quad\quad MnSb_4Te_7\,\,\,(\,2\mid \,\,\,FM\,\,\mid c\,)}$ & \multicolumn{4}{c|}{1}& \multicolumn{5}{c}{0}\\ \hline
        $\mathrm{\quad\quad\quad MnSb_6Te_{10}\,(\,2\mid \,\,\,FM\,\,\mid a\,)}$ & \multicolumn{4}{c|}{0}& \multicolumn{5}{c}{1}\\ \hline
        $\mathrm{\quad\quad\quad MnBi_2Te_4\,\,\,\,(\,3\mid \,\,\,FM\,\,\mid a\,)}$  & \multicolumn{9}{c}{1} \\ \hline
        \hline
    \end{tabular}
\end{table}

\subsection{Band structures of three-layer $\rm{MnBi}_2 \rm{Te}_4$ with strain}
Band structures of three-layer $\rm{MnBi}_2 \rm{Te}_4$ with strain are shown in Fig.~\ref{fig:MBT_bandstrain}. The indirect gaps of three-layer $\rm{MnBi}_2 \rm{Te}_4$ (3L MBT) in the presence of in-plane strain -2\%, 0\% and 2\% are 29.7 meV, 46.3 meV and 53.0 meV, respectively. Larger gap results in better quantization of the calculated Chern number given the same $\mathbf{k}$ mesh.
\begin{figure}[h]
    \centering
    \includegraphics[width=0.8\textwidth]{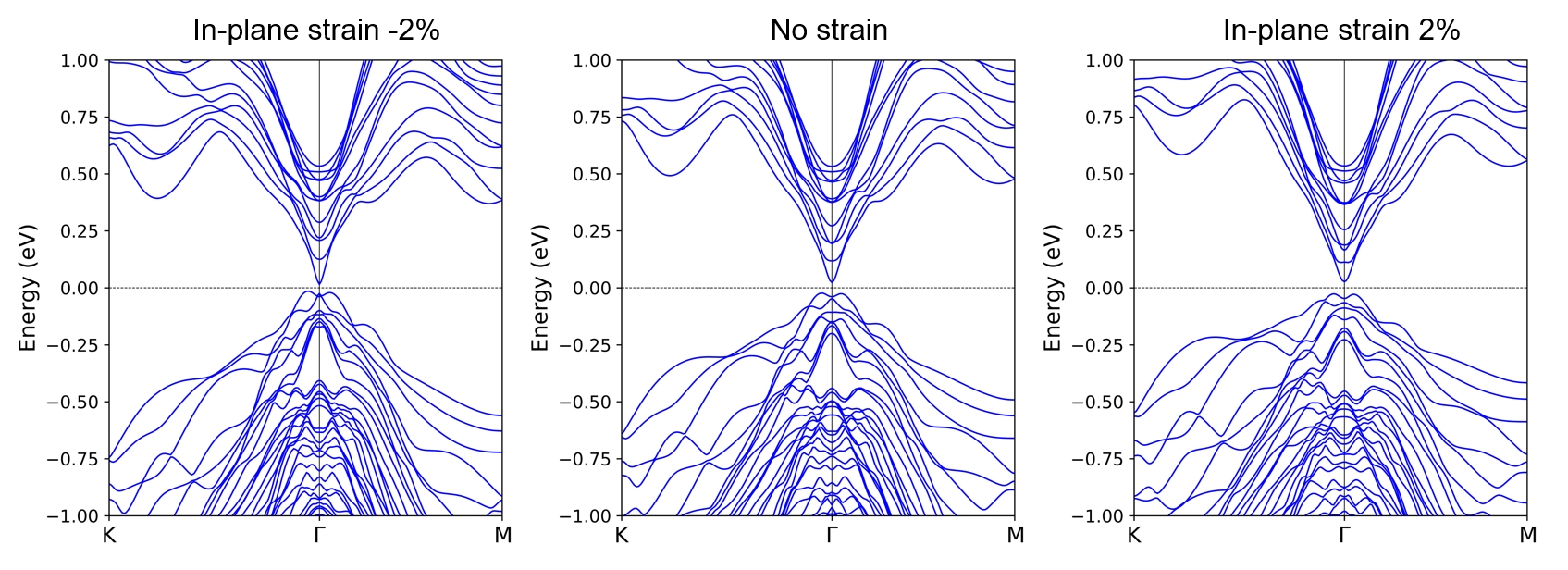}
    \caption{Band structures of 3L MBT in the presence of in-plane strain -2\%, 0\% and 2\% and the zero energy indicates chemical potential.}
    \label{fig:MBT_bandstrain}
\end{figure}

\section{Results using another expression of layer-resolved orbital magnetization}
Recently, Seleznev et al. \cite{seleznez-surfaceM-prb-2023} have shown that multiple formulations for the local marker of orbital magnetization exist. Those formulations are not equivalent for systems belonging to certain symmetry classes. The authors have identified one specific form of the marker as the most physically meaningful since it can consistently predict the correct hinge current. To validate our results, we derive the expression of layer-resolved orbital magnetization from the local marker proposed in \cite{seleznez-surfaceM-prb-2023}
\begin{equation}
    \label{eq:M_def_linear}
    M_z =  \frac{e}{3\hbar A}{\rm Im\,Tr}\{PxQHQyP-QxPHPyQ\} +  \frac{e}{3\hbar A}{\rm Im\,Tr}\{H(QyPxQ-PyQxP)\} +  \frac{e}{3\hbar A}{\rm Im\,Tr}\{(QyPxQ-PyQxP)H\},
\end{equation}
which is an equal weight linear combination of three possible operator orderings in the trace operation. 

The derivation of the definition layer-resolved orbital magnetization from Eq.~\eqref{eq:M_def_linear} follows the same algebra as outlined in Sec.~\I.B. We then check the validity of our proposed quantization rule using the new definition under exactly the same set-up described in the main text. Remarkably, the quantization rule remains valid under this new definition, as shown in the subsequent Tables~\ref{Table_layer_resolved_PBC_linear}, \ref{Table_diff_PBC_linear} and \ref{Table_substrate_linear}, which correspond to Tables~\I, \II\  and \IV\  in the main text, respectively.

\begin{table}%[H] add [H] placement to break table across pages
    \caption{Layer-resolved $\alpha_{zz}^{(l)}$ in units of $e^2/h$ obtained by numerical finite difference using the definition Eq.~\eqref{eq:M_def_linear} proposed in \cite{seleznez-surfaceM-prb-2023}. Same parameters are used as for Table~\ref{tab:layer_resolved_PBC} and Table~\I\  in the main text.}
    \label{Table_layer_resolved_PBC_linear}
    \begin{ruledtabular}
    \begin{tabular}{c| c c c c c c}
    Layer Index &1&2&3&4&5&6\\
    \hline
    AA,5&1.9924&0.9971&-0.0010&-0.9990&-1.9943&\diagbox[dir=NW]{}{}\\
    AB,5&1.9377&0.9525&-0.0448&-1.0430&-2.0282&\diagbox[dir=NW]{}{}\\
    AA,6&2.5397&1.5444&0.5464&-0.4519&-1.4502&-2.4448\\
    AB,6&2.4957&1.5086&0.5113&-0.4842&-1.4828&-2.4680\\
    \end{tabular}
    \end{ruledtabular}
\end{table} 

\begin{table}%[H] add [H] placement to break table across pages
    \caption{Difference between $\alpha_{zz}^{(l)}$ of neighboring layers in units of $e^2/h$ obtained by numerical finite difference using the definition Eq.~\eqref{eq:M_def_linear} proposed in \cite{seleznez-surfaceM-prb-2023}. Same parameters are used as for Table~\ref{tab:layer_resolved_PBC} and Table~\II\  in the main text.}
    \label{Table_diff_PBC_linear}
    \begin{ruledtabular}
    \begin{tabular}{c| c c c c c c}
    $\alpha_{zz}^{(l+1)}-\alpha_{zz}^{(l)}$ &$l=1$&$l=2$&$l=3$&$l=4$&$l=5$\\
    \hline
    AA,5&-0.9953&-0.9982&-0.9980&-0.9953&\diagbox[dir=NW]{}{}\\
    AB,5&-0.9852&-0.9974&-0.9982&-0.9852&\diagbox[dir=NW]{}{}\\
    AA,6&-0.9953&-0.9980&-0.9983&-0.9983&-0.9946\\
    AB,6&-0.9871&-0.9973&-0.9955&-0.9986&-0.9852\\
    \end{tabular}
    \end{ruledtabular}
\end{table}

\begin{table}
    \centering
    \caption{Difference between neighboring layer-resolved response in units of $e^2/h$ in systems with substrate using the definition Eq.~\eqref{eq:M_def_linear}. Same parameters are used as for Table~\IV\  in the main text.}
    \label{Table_substrate_linear}
    \begin{ruledtabular}
    \begin{tabular}{c| c c c c}
    $\alpha_{zz}^{(l+1)}-\alpha_{zz}^{(l)}$ & $l=1$ & $l=2$ & $l=3$ & $l=4$ \\
    \hline
    $t_3=0,t_4=0$ & -0.9972&-0.9972&-0.9972&-0.9972\\
    % \cline{2-2}\cline{5-9}
    $t_3=1/6,t_4=0$ & -0.9972&-0.9972&-0.9972&-0.9972\\
    % \cline{2-2}\cline{5-9}
    $t_3=0.3,t_4=0$ & -0.9972&-0.9972&-0.9972&-0.9972\\
    % \cline{2-2}\cline{4-9}
    $t_3=1/6,t_4=3t_3/4$ & -0.9954&-0.9978&-0.9978&-0.9955\\
    % \cline{2-2}\cline{5-9}
    $t_3=0.3,t_4=3t_3/4$ & -0.9910&-0.9991&-0.9990&-0.9913\\
    \end{tabular}
    \end{ruledtabular}
\end{table}

\end{widetext}

\end{document}